\documentclass[conference]{IEEEtran}

\ifCLASSINFOpdf
\else
\fi

\usepackage{tikz}
\usepackage{amsmath}
\usepackage{tabularx}
\usepackage{filecontents}
\usepackage{graphicx}
\usepackage{textcomp}
\usepackage{xcolor}
\usepackage{epsfig}
\usepackage{endnotes}
\usepackage{tikz}
\usepackage{amsmath}
\usepackage{amssymb}
\usepackage{amsfonts}
\usepackage{filecontents}
\usepackage{multirow}
\usepackage{booktabs}
\usepackage[mathscr]{euscript}
\usepackage{url}
\usepackage{xspace}
\usepackage{siunitx}
\usepackage{mathtools}
\usepackage{algorithm}
\usepackage{algorithmicx}
\usepackage[noend]{algpseudocode}
\usepackage{balance}
\usepackage{makecell}
\usepackage{comment}
\usepackage{wasysym}
\usepackage{pifont}
\usepackage{hhline}
\usepackage{hyperref}
\usepackage{extarrows}
\usepackage{colortbl} 
\usepackage{enumitem}
\usepackage[normalem]{ulem}
\usepackage{tcolorbox}

\usepackage{rotating}
\usepackage{tabularray}
\usepackage{makecell}
\usepackage{circledsteps}
\usepackage{diagbox}

\def\BibTeX{{\rm B\kern-.05em{\sc i\kern-.025em b}\kern-.08em
    T\kern-.1667em\lower.7ex\hbox{E}\kern-.125emX}}

\newcolumntype{C}[1]{>{\centering\let\newline\\\arraybackslash\hspace{0pt}}m{#1}}
\DeclareMathAlphabet{\mathbfit}{OML}{cmm}{b}{it}

\newcommand{\MR}[1]{{}}

\newcommand{\as}[1]{\textcolor{red}{}} 

\newcommand{\eqp}{\begingroup\hypersetup{hidelinks}\hyperref[subsec:Hardware]{\mbox{$\mathcal{E}_q$}}\endgroup}
\newcommand{\ax}{\begingroup\hypersetup{hidelinks}\hyperref[subsec:BusAccess]{\mbox{$\mathcal{A}_c$}}\endgroup}
\newcommand{\abl}{\begingroup\hypersetup{hidelinks}\hyperref[subsec:AttackAssessment]{\mbox{$\mathcal{A}_m$}}\endgroup}
\newcommand{\prev}{\begingroup\hypersetup{hidelinks}\hyperref[subsec:AttackAssessment]{\mbox{$\mathcal{V}_r$}}\endgroup}
\newcommand{\diff}{\begingroup\hypersetup{hidelinks}\hyperref[subsec:AttackAssessment]{\mbox{$\mathcal{D}_f$}}\endgroup}
\newcommand{\ins}{\begingroup\hypersetup{hidelinks}\hyperref[subsec:DefenseAssessment]{\mbox{$\mathcal{I}_r$}}\endgroup}
\newcommand{\cov}{\begingroup\hypersetup{hidelinks}\hyperref[subsec:DefenseAssessment]{\mbox{$\mathcal{C}_v$}}\endgroup}
\newcommand{\ecu}{\begingroup\hypersetup{hidelinks}\hyperref[subsec:DefenseAssessment]{\mbox{$\mathcal{E}_e$}}\endgroup}
\newcommand{\bus}{\begingroup\hypersetup{hidelinks}\hyperref[subsec:DefenseAssessment]{\mbox{$\mathcal{B}_p$}}\endgroup}
\newcommand{\infra}{\begingroup\hypersetup{hidelinks}\hyperref[subsec:DefenseAssessment]{\mbox{$\mathcal{I}_c$}}\endgroup}
\newcommand{\val}{\begingroup\hypersetup{hidelinks}\hyperref[subsec:DefenseAssessment]{\mbox{$\mathcal{V}_l$}}\endgroup}

\newcommand{\standard}{\Circle}
\newcommand{\widespread}{\LEFTcircle}
\newcommand{\specific}{\CIRCLE}
\newcommand{\low}{\Circle}
\newcommand{\moderate}{\LEFTcircle}
\newcommand{\high}{\CIRCLE}
\newcommand{\Ca}{\begingroup\hypersetup{hidelinks}\hyperref[subsubsec:reveng]{\mbox{C1}}\endgroup}
\newcommand{\Cb}{\begingroup\hypersetup{hidelinks}\hyperref[subsubsec:leakage]{\mbox{C2}}\endgroup}
\newcommand{\Cc}{\begingroup\hypersetup{hidelinks}\hyperref[subsubsec:errorRecon]{\mbox{C3}}\endgroup}
\newcommand{\Cd}{\begingroup\hypersetup{hidelinks}\hyperref[subsubsec:topology]{\mbox{C4}}\endgroup}
\newcommand{\Ce}{\begingroup\hypersetup{hidelinks}\hyperref[subsubsec:electrosnif]{\mbox{C5}}\endgroup}

\newcommand{\Ia}{\begingroup\hypersetup{hidelinks}\hyperref[subsubsec:datafabricate]{\mbox{I1}}\endgroup}
\newcommand{\Ib}{\begingroup\hypersetup{hidelinks}\hyperref[subsubsec:memoryManip]{\mbox{I2}}\endgroup}
\newcommand{\Ic}{\begingroup\hypersetup{hidelinks}\hyperref[subsubsec:bypassUDSAuth]{\mbox{I3}}\endgroup}
\newcommand{\Id}{\begingroup\hypersetup{hidelinks}\hyperref[subsubsec:impersonation]{\mbox{I4}}\endgroup}
\newcommand{\Ie}{\begingroup\hypersetup{hidelinks}\hyperref[subsubsec:frameHijack]{\mbox{I5}}\endgroup}
\newcommand{\IF}{\begingroup\hypersetup{hidelinks}\hyperref[subsubsec:dblrcv]{\mbox{I6}}\endgroup}
\newcommand{\Ig}{\begingroup\hypersetup{hidelinks}\hyperref[subsubsec:unOrth]{\mbox{I7}}\endgroup}
\newcommand{\Ih}{\begingroup\hypersetup{hidelinks}\hyperref[subsubsec:polysemantic]{\mbox{I8}}\endgroup}
\newcommand{\Ii}{\begingroup\hypersetup{hidelinks}\hyperref[subsubsec:frmtmpr]{\mbox{I9}}\endgroup}
\newcommand{\Ij}{\begingroup\hypersetup{hidelinks}\hyperref[subsubsec:volCorr]{\mbox{I10}}\endgroup}

\newcommand{\Aa}{\begingroup\hypersetup{hidelinks}\hyperref[subsubsec:ECUDisable]{\mbox{A1}}\endgroup}
\newcommand{\Ab}{\begingroup\hypersetup{hidelinks}\hyperref[subsubsec:j1939ConExhaust]{\mbox{A2}}\endgroup}
\newcommand{\Ac}{\begingroup\hypersetup{hidelinks}\hyperref[subsubsec:ErrorInj]{\mbox{A3}}\endgroup}
\newcommand{\Ad}{\begingroup\hypersetup{hidelinks}\hyperref[subsubsec:BusOff]{\mbox{A4}}\endgroup}
\newcommand{\Ae}{\begingroup\hypersetup{hidelinks}\hyperref[subsubsec:flooding]{\mbox{A5}}\endgroup}
\newcommand{\Af}{\begingroup\hypersetup{hidelinks}\hyperref[subsubsec:synchDis]{\mbox{A6}}\endgroup}
\newcommand{\Ag}{\begingroup\hypersetup{hidelinks}\hyperref[subsubsec:sigAttenuate]{\mbox{A7}}\endgroup}

\newcommand{\DCa}{\begingroup\hypersetup{hidelinks}\hyperref[subsubsec:payloadObf]{\mbox{DC1}}\endgroup}
\newcommand{\DCb}{\begingroup\hypersetup{hidelinks}\hyperref[subsubsec:IDObf]{\mbox{DC2}}\endgroup}
\newcommand{\DCc}{\begingroup\hypersetup{hidelinks}\hyperref[subsubsec:shielding]{\mbox{DC3}}\endgroup}

\newcommand{\DIa}{\begingroup\hypersetup{hidelinks}\hyperref[subsubsec:msgAuth]{\mbox{DI4}}\endgroup}
\newcommand{\DIb}{\begingroup\hypersetup{hidelinks}\hyperref[subsubsec:content_inspect]{\mbox{DI1}}\endgroup}
\newcommand{\DIc}{\begingroup\hypersetup{hidelinks}\hyperref[subsubsec:ProtocolCheck]{\mbox{DI2}}\endgroup}
\newcommand{\DId}{\begingroup\hypersetup{hidelinks}\hyperref[subsubsec:ecu_hardening]{\mbox{DI3}}\endgroup}
\newcommand{\DIe}{\begingroup\hypersetup{hidelinks}\hyperref[subsubsec:FrameDest]{\mbox{DI5}}\endgroup}
\newcommand{\DIF}{\begingroup\hypersetup{hidelinks}\hyperref[subsubsec:victimRes]{\mbox{DI6}}\endgroup}
\newcommand{\DIg}{\begingroup\hypersetup{hidelinks}\hyperref[subsubsec:unify_sp]{\mbox{DI8}}\endgroup}
\newcommand{\DIh}{\begingroup\hypersetup{hidelinks}\hyperref[subsubsec:zbcan]{\mbox{DI9}}\endgroup}
\newcommand{\DIi}{\begingroup\hypersetup{hidelinks}\hyperref[subsubsec:IDS]{\mbox{DI7}}\endgroup}
\newcommand{\DIj}{\begingroup\hypersetup{hidelinks}\hyperref[subsubsec:layerIIIDS]{\mbox{DI10}}\endgroup}
\newcommand{\DIk}{\begingroup\hypersetup{hidelinks}\hyperref[subsubsec:guardian]{\mbox{DI11}}\endgroup}

\newcommand{\DAa}{\begingroup\hypersetup{hidelinks}\hyperref[subsubsec:id_prio_dis]{\mbox{DA3}}\endgroup}
\newcommand{\DAb}{\begingroup\hypersetup{hidelinks}\hyperref[subsubsec:oneShot]{\mbox{DA4}}\endgroup}
\newcommand{\DAc}{\begingroup\hypersetup{hidelinks}\hyperref[subsubsec:err_ids]{\mbox{DA1}}\endgroup}
\newcommand{\DAd}{\begingroup\hypersetup{hidelinks}\hyperref[subsubsec:nodeSuspend]{\mbox{DA2}}\endgroup}
\newcommand{\DAe}{\begingroup\hypersetup{hidelinks}\hyperref[subsubsec:fragmentation]{\mbox{DA5}}\endgroup}

\usepackage{microtype}
\usepackage{setspace}

\newcommand{\shortsectionBf}[1]{\noindent {\bf #1}}

\newcommand{\shortsectionEmph}[1]{\noindent {{\tiny $\blacksquare$~~}\em #1}}

\newcounter{paranum}[subsection]           
 
\newcommand{\mynum}{\arabic{paranum}} 

\newcommand{\mysub}[1]{\vspace{2pt}\noindent\refstepcounter{paranum}{\bf #1}\vspace{2pt}}

\newcommand{\mysubsub}[1]{\noindent\refstepcounter{paranum}{\bf\mynum. #1.}}

\newcommand{\confsub}[1]{\noindent\refstepcounter{paranum}{\bf C\mynum. #1.}}

\newcommand{\intsub}[1]{\noindent\refstepcounter{paranum}{\bf I\mynum. #1.}}

\newcommand{\avsub}[1]{\noindent\refstepcounter{paranum}{\bf A\mynum. #1.}}

\newcommand{\Dconfsub}[1]{\noindent\refstepcounter{paranum}{\bf DC\mynum. #1.}}

\newcommand{\Dintsub}[1]{\noindent\refstepcounter{paranum}{\bf DI\mynum. #1.}}

\newcommand{\Davsub}[1]{\noindent\refstepcounter{paranum}{\bf DA\mynum. #1.}}

\begin{document}

\title{SoK: Kicking CAN Down the Road \\Systematizing CAN Security Knowledge}

\author{\IEEEauthorblockN{Khaled Serag}
	\IEEEauthorblockA{Qatar Computing Research Institute\\
		kseragalsharif@hbku.edu.qa}
	\and
	\IEEEauthorblockN{Zhaozhou Tang}
	\IEEEauthorblockA{Georgia Institute of Technology\\
		ztang306@gatech.edu}
	\and
	\IEEEauthorblockN{Sungwoo Kim}
	\IEEEauthorblockA{Purdue University\\
		kim3583@purdue.edu}
    \and\IEEEauthorblockN{Vireshwar Kumar}
	\IEEEauthorblockA{IIT Delhi\\
		viresh@cse.iitd.ac.in}
	\and
	\IEEEauthorblockN{Dave (Jing) Tian}
	\IEEEauthorblockA{Purdue University\\
		daveti@purdue.edu}
	\and
	\IEEEauthorblockN{Saman Zonouz}
	\IEEEauthorblockA{Georgia Institute of Technology\\
		szonouz6@gatech.edu}
    \and
    \IEEEauthorblockN{Raheem Beyah}
	\IEEEauthorblockA{Georgia Institute of Technology\\
		rbeyah@coe.gatech.edu}
	\and
	\IEEEauthorblockN{Z. Berkay Celik}
	\IEEEauthorblockA{Purdue University\\
		zcelik@purdue.edu}
	\and
	\IEEEauthorblockN{Dongyan Xu}
	\IEEEauthorblockA{Purdue University\\
		dxu@purdue.edu}}

\IEEEoverridecommandlockouts
\makeatletter\def\@IEEEpubidpullup{6.5\baselineskip}\makeatother
\IEEEpubid{\parbox{\columnwidth}{
		Network and Distributed System Security (NDSS) Symposium 2027\\
		22--26 March 2027, Seoul, Republic of Korea\\
		ISBN 978-1-970672-09-1\\  
		https://dx.doi.org/10.14722/ndss.2027.230993\\
		www.ndss-symposium.org
}
\hspace{\columnsep}\makebox[\columnwidth]{}}

\maketitle

\begin{abstract}

For decades, the Controller Area Network (CAN) has served as the primary in-vehicle bus (IVB), extending its use to many non-vehicular systems. In recent years, CAN security has been intensively scrutinized, yielding extensive research literature. Despite its wealth, the literature lacks structured systematization, complicating efforts to assess and compare attack severity, defense efficacy, security gaps, and root causes. This leaves many defenders uncertain about the relevance of specific attacks or defenses to their systems, and even whether CAN’s security problems are truly CAN-specific. \MR{9}As  newer IVBs emerge, this matters beyond CAN: if CAN’s root causes are not CAN-specific, replacing CAN may only move its problems to a new standard.

In this paper, we systematize CAN security knowledge, presenting a comprehensive taxonomy and assessment models of attackers, attacks, and defenses. We identify replicable attacks and defense gaps, and investigate their root causes to determine their exclusivity to the CAN standard. We then investigate whether those root causes appear in three emerging IVBs and assess their effectiveness in solving fundamental CAN security problems. Our findings challenge common perceptions: CAN is more securable than perceived, most of its insecurity root causes are shared across IVBs, and merely adopting newer IVB technology does not solve persistent security problems. We conclude by suggesting that securing future in-vehicle communication requires addressing shared root causes, and we propose four research directions with the most promising potential.

\end{abstract}
\IEEEpeerreviewmaketitle

\section{Introduction}
\label{sec:introduction}

In the 1970s, integrated circuits began appearing in vehicles as Electronic Control Units (ECUs) to digitize certain functions. By the 1980s, vehicles contained numerous ECUs, but point-to-point connections introduced significant wiring complexity. In 1986, Bosch released the Controller Area Network (CAN) standard, tackling issues including wiring, interference resistance, and decentralization, enabling modern vehicles to feature dozens of ECUs, coordinating essential functions. Today, CAN is present in all vehicles and has extended its use to industrial automation, avionics, medical devices, and many other arenas.

Designed in an era of network isolation, increasing connectivity has exposed CAN to many exploits, triggering a research race between attack and defense, and resulting in very rich and diverse literature. While rich in technical detail, it predominantly emphasizes technical novelty and how each paper differs from closely related ones—often making papers pursuing the same end goal look unrelated. As a result, the broader picture remains diffuse, especially from a defense standpoint. A defender starts from a priority-ranked hierarchy of security objectives (e.g., availability of a safety function, integrity of a particular sensor value, or confidentiality of certain data) their system must preserve, assuming a particular threat model. They need a way to turn the vast literature into something usable: which attack papers target the same security objective, what is replicable on their systems, what threat models they presuppose, what performance impact a defense entails, and what defense gaps they should know about.

\begin{figure*}[t]
  \centering
  \includegraphics[width=\linewidth]{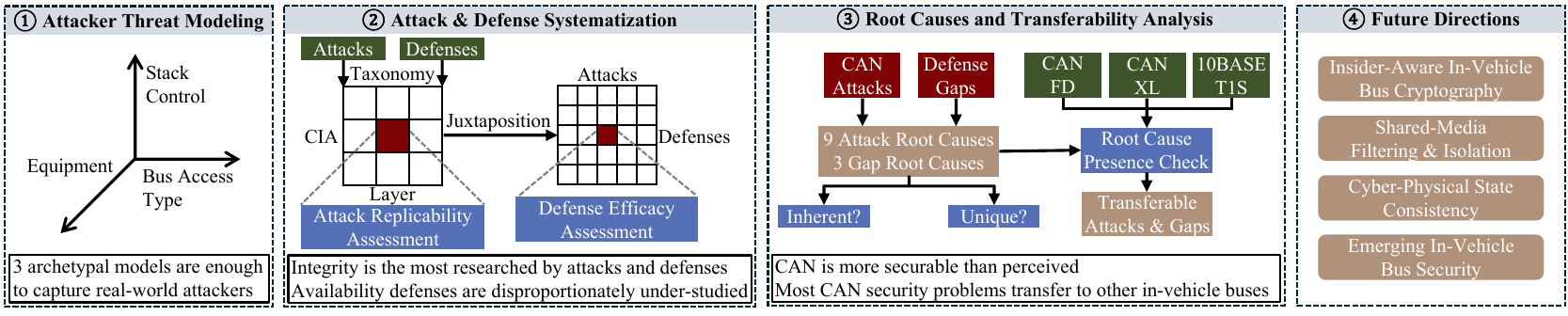}

  \caption{Overview of the systematization flow and the paper's main contributions.}
  \label{fig:summary}

\end{figure*}

In the absence of a systematization framework that could help with these questions, the wealth of CAN security literature has paradoxically worked to its disadvantage. This may cultivate a perception of CAN as hopelessly insecurable, especially as some voices explicitly call for replacing CAN with newer in-vehicle bus (IVB) technologies, including CAN evolutions and alternative buses, with security among the primary stated motivations \cite{kickCAN1,kickCAN2}. Several valuable surveys provided nascent systematization attempts \cite{jo2021survey, loto2021survey, avatefipour2018state}, some focusing on classifying attackers based on attack surfaces and entry points \cite{rathore2022vehicle, bozdal2020evaluation, aliwa2021cyberattacks}, and some on specific defense approaches \cite{rajapaksha2023ai,wu2019survey} or authentication protocols \cite{lotto2024survey}. Nonetheless, a coherent and comprehensive systematization framework remains conspicuously absent. Without a framework to articulate attacker privileges, map privileges to attacks, group attacks with similar security objectives, and assess the significance of an attack or a defense, it is infeasible to identify replicable attacks, defense gaps, or root causes. Finally, no prior research systematically analyzed whether CAN security problems are exclusive to the CAN standard, or whether they stem from broader structural limitations of in-vehicle communication. As interest in new IVB technologies grows, the issue becomes more pressing. Misunderstandings about CAN’s core security issues may foster the belief that simply adopting newer technology resolves CAN security issues, unknowingly reproducing them in other IVBs.

This paper systematizes existing CAN security knowledge, presenting a comprehensive defender-oriented systematization framework. \MR{3}We take this perspective because OEMs, practitioners, researchers, and standards bodies are, to varying extents, all defenders seeking to make informed decisions. Our systematization serves each directly. OEMs can anticipate which security problems transfer when adopting a new technology. Practitioners can determine which attacks are possible in their systems and which defenses address them. Researchers can focus their efforts on gaps instead of already crowded areas. Standards bodies can identify the root causes underlying security weaknesses in standards. 

As shown in Fig. \ref{fig:summary}, we start with multi-dimensional attacker modeling, contextualizing ability, access, and equipment, to understand the various attacker models appearing in the literature. Second, we provide an objective-based taxonomy and assessment model for the attacks. The goal is to understand the various attack objectives appearing in the literature, the different variants, and their replicability across CAN systems in terms of their difficulty and required attacker models or implementation flaws. We do the same for defenses, taxonomizing the existing defense landscape and assessing their effectiveness in tackling the taxonomized attacks in terms of security, infrastructure, and performance impact. Juxtaposing attacks and defenses aims to expose what is most targeted, what is most defended, which attacks have defense gaps, and where research is concentrated or lacking. Third, building on these outcomes, we then analyze the root causes behind CAN attacks and defense gaps to understand the extent to which they are inherent or unique to CAN. We then conduct an investigation of three representative emerging IVBs: two CAN evolutions, CAN FD and CAN XL, and one alternative bus, 10BASE-T1S. The goal is to determine the extent to which CAN problems are transferable to other IVBs and the extent to which these IVBs solve CAN security issues. Finally, we identify four research directions with the highest potential to improve the security of in-vehicle communication.

Our study results in interesting outcomes. \MR{3\&9}Relative to threat models, our multi-dimensional analysis reveals that the seemingly diverse attacker models used across the literature could collapse into only three attacker models that are enough to adequately represent the various models in real life. This reduction is small enough to be actionable while preserving the distinctions that matter for security. It allows attacks to be compared under common attacker models and gives practitioners a direct mapping from attacker privileges to feasible attacks and applicable defenses. It also exposes cases where a defense appears effective only because it assumes a weaker attacker. In terms of research interest, while integrity receives the most attention in the attack literature, it also receives the highest attention in the defense literature, yielding several effective defenses. On the other hand, availability appears to be the most exposed CIA element, with the highest proportion of highly replicable attacks with no effective defenses and the least amount of research papers. In terms of root-cause and transferability analysis, we reveal that CAN is more securable than perceived, that most CAN insecurity problems are rooted in causes inherent to all single-channel communication media or vehicular systems, that most of them are transferable to other IVBs, and that none of its defense gaps are rooted in causes unique to CAN. Finally, we suggest that securing CAN evolutions, and all future in-vehicle communication down the road, requires addressing shared root causes more than CAN-specific causes. Overall, we make the following contributions:

\setlist{nosep}
\begin{itemize} [leftmargin=* ]
    \item We develop a defender-oriented CAN security systematization framework, including a CAN-oriented abstraction stack, a multi-dimensional attacker model, and assessment criteria for attacks and defenses.
   
    \item We survey, taxonomize, and assess CAN attacks and defenses, covering attack goals, attack replicability, attack variants, defense approaches, and defense effectiveness.
   
    \item We juxtapose attacks and defenses to identify highly replicable attacks, expose defense gaps, and show where research is concentrated or lacking.
   
    \item We analyze the root causes behind CAN attacks and defense gaps to determine the extent to which they are inherent or unique to the CAN standard.
   
    \item We investigate whether CAN 2.0 security problems transfer to other IVBs through a targeted investigation of three representative emerging IVBs: CAN FD, CAN XL, and 10BASE-T1S.\looseness-1

    \item We identify four research directions with the highest potential to improve in-vehicle communication security.

\end{itemize}

\begin{figure*}[t]
  \centering
  \includegraphics[width=\linewidth]{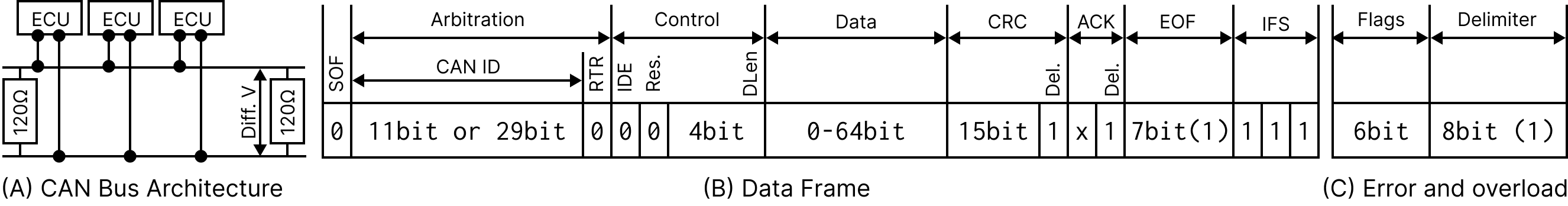}

  \caption{CAN bus architecture, data frame, and error \& overload frame formats.}
  \label{fig:frameform}
  \label{fig:errorform}

\end{figure*}

\section{CAN Basics}
\label{sec:background}





     



\shortsectionBf{CAN Operation.} Fig. \ref{fig:frameform}-A illustrates the masterless, publish-subscribe nature of the CAN bus, consisting of two wires terminated by resistors. Their voltage difference represents bits: zeros (dominant) and ones (recessive). Nodes may initiate transmission when the bus is idle, transmitting data bit-by-bit. Fig. \ref{fig:frameform}-B illustrates the format of a data frame. If two nodes transmit simultaneously, collision is resolved using arbitration over the ID and RTR fields: when one node sends a 0 and another a 1, the node transmitting 1 halts, yielding to the 0 transmitter. Lower IDs thus have higher priority. While nodes may transmit multiple message IDs, each ID is typically associated with only one ECU. CAN frames begin with a Start of Frame (SoF) bit (logical 0) and conclude with 7 End-of-Frame (EoF) bits (logical 1). A minimum Interframe Space (IFS) of 3 bits separates consecutive frames on the bus.

\shortsectionBf{Error and Overload Handling.}
\label{subsubsec:errors}
\label{subsubsec:errorStates} CAN defines 5 error types. Nodes update a transmit (TEC) and a receive error counter (REC). Transmit errors increase the TEC by 8. Others increase REC. If either counter exceeds 127, nodes enter the error-passive state with stricter transmission rules. If transmit errors continue (TEC $>$ 255), they move into the bus-off state and stop participating in communication. Nodes signal errors using an error frame composed of a flag and a delimiter (Fig. \ref{fig:errorform}-C). In the error active state, the flag is dominant, interrupting ongoing transmission. In the error passive state, it is recessive, invisible, and cannot interrupt transmission. Overload frames have a similar format to active errors, but do not increase error counters and occur only between frames by resource-constrained nodes to slow down bus communication.\looseness-1

\shortsectionBf{CAN Stack Model.}
\label{subsubsec:Netwok_Model} CAN aligns with the two bottom OSI layers. To communicate, a typical ECU (Fig. \ref{fig:ECUArch}) uses a CAN controller (enforces protocol rules) connected to a transceiver (converts 0s and 1s to differential voltage). Although the functionalities of the controller and transceiver highly align with OSI's data-link and physical layers, the alignment is not exact. Consequently, we choose to model CAN stack after the typical components of an ECU to better fit CAN's nature:\looseness-1  

\shortsectionEmph{Transwire Layer:} Encompassing the transceiver and wires, it roughly corresponds with the OSI's physical layer.

\shortsectionEmph{Controller Layer:} Encompassing only the CAN controller, this layer roughly corresponds with the OSI's data-link layer.

\shortsectionEmph{Higher Layer:} Higher-layer protocols that extend CAN functionality to higher OSI layers (e.g., transport, session, etc.) are not part of the CAN standard, but are part of the CAN stack. We consider them collectively as the ``higher layer", encompassing all that interacts with the controller from above.

\shortsectionBf{Common Higher-Layer Protocols.} Many protocols can run on top of CAN. Two common examples are UDS and J1939. \textit{1) Unified Diagnostic Services (UDS):} a client-server diagnostic protocol in which a client, typically a diagnostic tool connected through the OBD-II port, communicates with ECUs acting as servers. UDS supports privileged operations such as reading or modifying ECU memory and firmware, and disabling ECU functionalities. These operations can be protected using seed-key authentication. \textit{2) J1939:} a protocol widely used for communication and diagnostics in trucks. It supports broadcast and request-response communication, packet fragmentation across multiple CAN frames for payloads exceeding 8 bytes, and dynamic higher-layer source-address assignment through an address-claim procedure after network initialization.



\section{Methodology \& Scope}
\label{sec:methodology}

\label{subsec:scope}

\label{subsec:TaxonomyandAss}
\label{subsubsec:taxonomy}

\shortsectionBf{Taxonomy.} To group attack papers that appear disparate in the literature by their underlying attack objective, we first place the paper in a coarse target bin defined by the intersection of the CIA triad and CAN stack layer (Sec.~\ref{sec:background}) for each attack a paper proposes. A paper may thus appear in multiple bins if it proposes attacks against different targets. Within each bin, we identify the specific attack objectives and group papers pursuing the same objective into one attack category, even when their execution paths differ. This groups variants of the same attack goal and separates attacks rooted in lower-layer standard properties from those exploiting higher-layer protocols or implementations, which later supports root-cause analysis.

Our defense taxonomy uses the same CIA $\times$ layer grid to enable direct juxtaposition with attacks. Since a defense may cover several properties, we taxonomize each defense idea by its primary protected CIA property and the deepest protected layer within that property, while noting protection against other objectives when relevant. A paper proposing multiple defense ideas may appear in multiple bins. This exposes misleading coverage, such as higher-layer integrity protection masking a controller-layer integrity weakness, and reveals where defenses are crowded, missing, or ineffective.

\shortsectionBf{Inclusion \& Exclusion Criteria.} \MR{1}Our systematization covers security knowledge on CAN 2.0. We include attacks launched by devices directly interfacing with the bus and targeting the confidentiality, integrity, or availability of the bus, its nodes, or the system it operates within (e.g., a vehicle). We include defenses developed for these attacks, and some CAN-adaptable solutions proposed for non-CAN systems. We exclude attacks launched by non-bus-interfacing devices, such as attacks on external vehicle sensors by a non-bus-connected device.\looseness-1

\shortsectionBf{Paper Selection Methodology.} \MR{1}We followed a two-stage approach consisting of an initial database search followed by manual citation-graph traversal. We searched for papers published between 1983, when Bosch began developing CAN, and June 1, 2025 in eight major security conferences: NDSS, IEEE S\&P, USENIX, CCS, ACSAC, ASIACCS, RAID, and EuroS\&P, using the keywords “vehicle” and “Controller Area Network.” This yielded 123 papers, of which 28 met our inclusion criteria and formed the initial seed. We then traversed their cited and citing papers without restricting venue or type, including relevant industry white papers and technical reports.

As the citation traversal expanded beyond the initial venues and across four decades of research, the candidate literature grew substantially, encompassing numerous works from specialized and smaller venues, many of which repurposed established attack and defense ideas with minor variations. We therefore retained only papers introducing a distinct attack or defense according to our taxonomy. Specifically, papers with the same bin, purpose, and assessment were treated as variations of the same idea, in which case we retained the earliest publication. We repeated the traversal until no additional eligible papers were identified, resulting in 150 papers.


\shortsectionBf{Paper Category Breakdown.} \MR{1\&2}Of the 150 papers, 43 proposed attacks only, 100 defenses only, and 7 both. Of the 100 defense-only papers, 17 were dual-use, meaning their techniques could also be used by attackers and were therefore additionally included in the attack taxonomy. In total, 67 papers contributed to attacks and 107 to defenses. Among attack papers, 33 targeted confidentiality, 28 integrity, and 23 availability, while 29 targeted the higher layer, 34 the controller layer, and 21 the transwire layer. Excluding dual-use papers, these counts were 17, 28, and 22 by security element and 29, 32, and 6 by layer, respectively. Among defense papers, 19 mainly addressed confidentiality, 88 integrity, and 5 availability, while 39 provided protection down to the higher layer, 70 the controller layer, and 17 the transwire layer. Note that counts do not sum to the corresponding paper totals because a paper may propose multiple distinct attacks or defenses that fall into different bins.\looseness-1 

Outside of the taxonomized papers, we additionally referenced relevant sources that do not propose novel CAN attacks or defenses. This included standard specification documents, review papers, news articles, user manuals and presentations related to the CAN ecosystem. We have provided a spreadsheet detailing each paper and its categorization at \cite{code}.

\section{Attacker Threat Modeling}
\label{sec:Attacker}

\begin{figure}[t]
  \centering
  \includegraphics[width=\columnwidth]{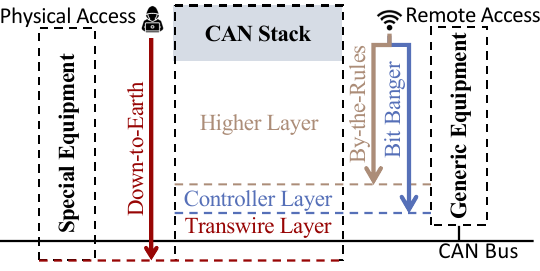}
  \caption{Typical ECU components, CAN stack, and archetypal attacker models.}
  \label{fig:models}
  \label{fig:ECUArch}
\end{figure}

Several surveys classify attackers based on bus access type only \cite{jo2021survey,rathore2022vehicle, bozdal2020evaluation, aliwa2021cyberattacks}. For a deeper understanding, we model attackers along three dimensions: \textit{bus access type} (Sec. \ref{subsec:BusAccess}), \textit{equipment} (Sec. \ref{subsec:Hardware}), and \textit{control} over CAN stack layers (Sec. \ref{subsec:Abilities}). A key revelation of this modeling is that, rather than producing a large number of attacker types, CAN attackers can be effectively reduced to only three archetypal models, discussed in Sec. \ref{subsec:Models}, that we use throughout the paper.

\subsection{Modeling Dimensions}\label{subsec:dimension}

\mysubsub{Bus \underline{Ac}cess Type (\ax)}\label{subsec:BusAccess}
Two levels appear in the literature:\looseness-1

\shortsectionEmph{\MR{5}Remote/On-ECU ($\Circle$):}
\label{subsubsec:RAccess} The weakest access type as the attacker has no physical access. \MR{5}The attacker here cannot attach any equipment and has to utilize the capabilities of an existing ECU remotely. Researchers showed this is achievable by compromising communication between smartphone apps and vehicles \cite{mazloom2016security,woo2014practical} or wireless OBD-II dongles \cite{wen2020plug}. Further, the growing connectivity of ECUs exposes new remote attack surfaces, such as Wi-Fi, Cellular, Bluetooth, remote keyless entry, TPMS, and radio \cite{miller2015remote, miller2014survey, checkoway2011comprehensive, nie2017free, nie2018ota}.

\shortsectionEmph{\MR{5}Physical/Side-ECU ($\CIRCLE$):}
\label{subsubsec:LAccess}
\label{subsubsec:FAccess}
The attacker has physical access to the bus that could be brief, sufficient to install a malicious device \MR{5}or a side ECU that may have wireless capabilities. Examples include accessing the OBD-II port during routine maintenance or service, or breaking through exposed areas such as behind the bumper or headlights \cite{BumperCAN}. More drastic changes, such as disconnecting ECUs or altering the bus topology require extensive access, which is difficult as it is often limited to authorized users.

\mysubsub{\underline{Eq}uipment (\eqp)}\label{subsec:Hardware} 
Two broad categories are used in attacks:

\shortsectionEmph{Generic ($\Circle$): }The standard off-the-shelf components found in mainstream ECUs. Any access type could use it but it is the only equipment type remote attackers can use as installing other equipment requires physical access. However, attackers may alter ECUs' logical connections through software, to deviate from the intended model (Fig. \ref{fig:ECUArch} and Sec. \ref{subsec:Abilities}).

\shortsectionEmph{Special ($\CIRCLE$): }
\label{subsubsec:SHardware}
The attacker designs their own equipment with its hardware layout fitted specifically for their intended attack. Physical access is required to install this equipment.

\mysubsub{\MR{9}Stack Control}\label{subsec:Abilities} The deepest layer of the CAN stack (Fig. \ref{fig:models}) the attacker could penetrate and control:

\shortsectionEmph{Higher Layer: }
\label{subsubsec:Hattacker} The attacker only controls the higher layer and has no control over the protocol controller or transceiver, except for what the higher-layer allows (e.g., baudrate). This control level is achievable using the weakest access type and equipment: remote access and generic equipment.


\shortsectionEmph{Controller Layer: }
\label{subsubsec:Dattacker} 
Attacker has full or partial control down to the controller layer, directly managing bus interactions. Traditionally, this required special hardware and physical access \cite{gessner2011designing, CANstress, serag2021exposing}. However, research \cite{kulandaivel2021cannon, de2022canflict, froschle2017analyzing} shows that it is  remotely attainable on many generic ECUs by exploiting peripheral clock gating to manipulate CAN frames mid-transmission, special controller modes, or pin conflicts to connect other peripherals, such as the SPI, I2C, or UART, directly to the CAN transceiver pins.

\shortsectionEmph{Transwire Layer:} Using {special} equipment and physical access, the attacker controls the entire CAN stack.

\subsection{Archetypal Attacker Models}
\label{subsec:Models}

\MR{9}The 3 outlined modeling dimensions contain 2, 2, and 3 possibilities, respectively. If simply multiplied, they would produce 12 combinations. However, most combinations are not meaningfully distinct because these dimensions are interdependent. The defining dimension is {stack control}: controlling a given CAN stack layer unlocks a specific set of {attack tool-blocks}, which form the building blocks used to launch the attacks in the literature. How control is achieved, through equipment or bus access type, affects feasibility rather than the resulting tool-block set. Moreover, control is cumulative: an attacker that controls a deeper layer inherits the tool-blocks of all higher layers. Thus, many apparent attacker types collapse into three equivalence classes defined by the deepest layer controlled. Under our stack model, this yields three archetypal attacker models (Fig. \ref{fig:models}), listed below.


\begin{figure}[t]
  \centering
  \includegraphics[width=\linewidth]{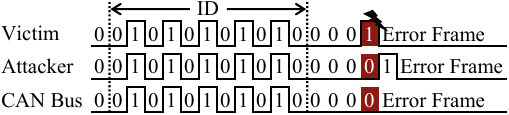}
  \caption{Inducing an error using the simultaneous transmission technique.}
  \label{fig:simTrans}
\end{figure}

\mysubsub{The By-the-Rules Attacker (\Circle)}
\label{subsubsec:BRules} As shown in Fig. \ref{fig:models}, this model is defined by its higher-layer control. It requires at least remote access and generic equipment. Due to its limited stack control, the attacker abides by all transwire and controller layer rules, confining the tool-blocks at their disposal to:\looseness-1

\shortsectionEmph{Frame Sniffing:} \label{subsubsec:frmsnf}
Reading messages assembled and processed by their uncompromised controller, without bit visibility. 

\shortsectionEmph{Frame Injection:} Injecting full data and remote request, but not error, overload, or malformed frames, only when the bus is idle through their intact controller and transceiver with full control over the data, ID, and DLC fields.

\shortsectionEmph{Simultaneous Transmission:}
\label{subsubsec:simultaneous}
This tool-block allows a node to bypass the collision resolution (arbitration) mechanism of CAN and transmit simultaneously with another node to achieve various purposes. It was discovered by Cho et al. \cite{cho2016error}. As shown in Fig. \ref{fig:simTrans}, if two nodes transmit two frames exactly at the same time with the same ID, both will transmit simultaneously until the first different bit, past the arbitration field. There, simultaneous transmission ends, resulting in an error frame (active or passive), and the termination of at least one transmission.\looseness-1

\mysubsub{The Bit-Banger (\LEFTcircle)}
\label{subsubsec:BBang} As shown in Fig. \ref{fig:models}, this model is defined by its control over the controller layer, requiring a minimum of remote access and generic equipment. The attacker here directly manages bus interactions typically handled by the CAN controller including encoding and rule enforcement. Attackers still adhere to transwire layer constraints (e.g., cannot change or read voltage levels, rise and fall times, etc.) as they do not control it. In addition to all the tool-blocks of by-the-rules attackers, it has this additional set:

\shortsectionEmph{Bit and Pulse Injection:} Attacker may inject individual bits or pulses. This may happen while the bus is idle or during another transmission, including performing 1$\rightarrow$0 bit-flips. 

\shortsectionEmph{Dynamic Bit-width and Baudrate Manipulation:} The attacker can dynamically control the width and baud of bits. While by-the-rules attackers could set their CAN controllers' baudrates to various rates, they cannot do it mid-transmission.

\shortsectionEmph{Bit and Pulse Sniffing:} Sniffing bits and pulses directly from the bus, including random pulses or individual bits transmitted during idle time, or within valid or invalid frames.

\mysubsub{The Down-to-Earth Attacker (\CIRCLE)}
\label{subsubsec:DEarth}  As shown in Fig. \ref{fig:models}, the attacker has full stack control down to the transwire layer, \MR{5}achievable only through physical access to attach special equipment or a side-ECU. In addition to the tool-blocks of the two previous models, it could manipulate all CAN rules (e.g., inject non-standard voltages, read exact voltage levels, rise and fall times, etc.).


\definecolor{Matrix}{rgb}{0.65,0.301,0.301}
\definecolor{Manhattan}{rgb}{0.7600,0.7955,0.8992}
\definecolor{BurntOrange}{rgb}{0.3373,0.4353,0.7216}
\definecolor{Bobo}{RGB}{176,145,122}

\newcolumntype{A}{>{\centering\arraybackslash}m{0.6cm}}

\newcommand{\atk}[1]{%
  \rotatebox[origin=c]{90}{\parbox{2.8cm}{\centering\small #1}}}

\begin{table*}[t]
\centering
\caption{CAN Attack Taxonomy and Assessment\\\textbf{HL}: Higher Layer,\quad \textbf{CT}: Controller Layer,\quad \textbf{TW}: Transwire Layer}
\label{tbl:attack_taxonomy}
\resizebox{\textwidth}{!}{%
\begin{tabular}{|m{3.2cm}|%
  >{\columncolor{Bobo}[\tabcolsep][\dimexpr\tabcolsep+\arrayrulewidth\relax]}A|A|>{\columncolor{Bobo}[\tabcolsep][\dimexpr\tabcolsep+\arrayrulewidth\relax]}A|A|A|
  >{\columncolor{Bobo}[\tabcolsep][\dimexpr\tabcolsep+\arrayrulewidth\relax]}A|A|A|
  >{\columncolor{Bobo}[\tabcolsep][\dimexpr\tabcolsep+\arrayrulewidth\relax]}A|>{\columncolor{Bobo}[\tabcolsep][\dimexpr\tabcolsep+\arrayrulewidth\relax]}A|
  >{\columncolor{Manhattan}[\tabcolsep][\dimexpr\tabcolsep+\arrayrulewidth\relax]}A|>{\columncolor{Manhattan}[\tabcolsep][\dimexpr\tabcolsep+\arrayrulewidth\relax]}A|A|A|
  >{\columncolor{Bobo}[\tabcolsep][\dimexpr\tabcolsep+\arrayrulewidth\relax]}A|
  A|A|
  >{\columncolor{Bobo}[\tabcolsep][\dimexpr\tabcolsep+\arrayrulewidth\relax]}A|>{\columncolor{Bobo}[\tabcolsep][\dimexpr\tabcolsep+\arrayrulewidth\relax]}A|>{\columncolor{Bobo}[\tabcolsep][\dimexpr\tabcolsep+\arrayrulewidth\relax]}A|>{\columncolor{Manhattan}[\tabcolsep][\dimexpr\tabcolsep+\arrayrulewidth\relax]}A|
  A|}
\hline
\textbf{CIA}
  & \multicolumn{5}{c|}{\textbf{Confidentiality}}
  & \multicolumn{10}{c|}{\textbf{Integrity}}
  & \multicolumn{7}{c|}{\textbf{Availability}}
  \\
\hline
\textbf{Layer}
  & \multicolumn{2}{c|}{\textbf{HL}}        
  & \multicolumn{1}{c|}{\textbf{CT}}        
  & \multicolumn{2}{c|}{\textbf{TW}}        
  & \multicolumn{3}{c|}{\textbf{HL}}        
  & \multicolumn{6}{c|}{\textbf{CT}}        
  & \multicolumn{1}{c|}{\textbf{TW}}        
  & \multicolumn{2}{c|}{\textbf{HL}}        
  & \multicolumn{4}{c|}{\textbf{CT}}        
  & \multicolumn{1}{c|}{\textbf{TW}}        
  \\
\hline
\textbf{Attack}
  & \atk{\Ca. Communication\ Surveillance}
  & \atk{\Cb. Memory Leakage}
  & \atk{\Cc. Network\ Reconnaissance}
  & \atk{\Cd. HW/Topology\ Profiling}
  & \atk{\Ce. EM Emissions}
  & \atk{\Ia. Fake Data Fabrication}
  & \atk{\Ib. Mem./FW Manipulation}
  & \atk{\Ic. Bypass Diag.\ Authentication}
  & \atk{\Id. Impersonation \& Replay}
  & \atk{\Ie. Frame Hijacking}
  & \atk{\IF. Double Receive}
  & \atk{\Ig. Unorthodox Frames}
  & \atk{\Ih. Poly-Semantic Frames}
  & \atk{\Ii. Frame Tampering}
  & \atk{\Ij. Physical FP Distortion}
  & \atk{\Aa. ECU Func.\ Disabling}
  & \atk{\Ab. Sybil Exhaustion}
  & \atk{\Ac. Error Injection}
  & \atk{\Ad. Error State Manipulation}
  & \atk{\Ae. Bus Access Denial}
  & \atk{\Af. Synchronization Disruption}
  & \atk{\Ag. Signal Attenuation}
  \\
\hline
\textbf{Attacker Model} (\abl)
  & $\Circle$              
  & $\Circle$              
  & $\Circle$              
  & $\CIRCLE$              
  & $\CIRCLE$              
  & $\Circle$              
  & $\Circle$              
  & $\Circle$              
  & $\Circle$              
  & $\Circle$              
  & $\LEFTcircle$          
  & $\LEFTcircle$          
  & $\LEFTcircle$          
  & $\LEFTcircle$          
  & $\Circle$              
  & $\Circle$              
  & $\Circle$              
  & $\Circle$              
  & $\Circle$              
  & $\Circle$              
  & $\LEFTcircle$          
  & $\LEFTcircle$          
  \\
\hline
\textbf{Vuln. Restrict.} (\prev)
  & $\Circle$              
  & $\CIRCLE$              
  & $\Circle$              
  & $\Circle$              
  & $\Circle$              
  & $\Circle$              
  & $\CIRCLE$              
  & $\CIRCLE$              
  & $\Circle$              
  & $\Circle$              
  & $\Circle$              
  & $\Circle$              
  & $\LEFTcircle$          
  & $\LEFTcircle$          
  & $\Circle$              
  & $\LEFTcircle$          
  & $\LEFTcircle$          
  & $\Circle$              
  & $\Circle$              
  & $\Circle$              
  & $\Circle$              
  & $\Circle$              
  \\
\hline
\textbf{Difficulty} (\diff)
  & $\LEFTcircle$          
  & $\Circle$              
  & $\Circle$              
  & $\LEFTcircle$          
  & $\LEFTcircle$          
  & $\LEFTcircle$          
  & $\Circle$              
  & $\LEFTcircle$          
  & $\Circle$              
  & $\Circle$              
  & $\Circle$              
  & $\Circle$              
  & $\LEFTcircle$          
  & $\LEFTcircle$          
  & $\Circle$              
  & $\Circle$              
  & $\Circle$              
  & $\Circle$              
  & $\Circle$              
  & $\Circle$              
  & $\Circle$              
  & $\CIRCLE$              
  \\
\hline
\end{tabular}%
}
\end{table*}

\section{Attacks On CAN}
\label{sec:Attacks}

\label{subsubsec:AttackCH}

In this section, we taxonomize CAN attack literature as described in Sec. \ref{sec:methodology}. We assess how likely these attacks are to recur across CAN-based systems by assessing their replicability beyond the original study setting using the criteria listed below. Tbl. \ref{tbl:attack_taxonomy} summarizes the resulting taxonomy, showing how attack targets distribute across CIA and layer, along with each attack's replicability assessment. In the table, highly replicable attacks are highlighted in light {brown} and {blue} for {by-the-rules} attackers and {bit-bangers}, respectively. \MR{2} We provide a spreadsheet at \cite{code} detailing the rationale for each assessment.

\shortsectionBf{Replicability Assessment Criteria.} \label{subsec:AttackAssessment}
Our goal is to help defense analysts judge whether a reported attack is likely to affect their CAN-based systems, enable uniform cross-paper comparison, and support later root-cause analysis. Therefore, we use {replicability} to assess CAN attacks: a system-agnostic measure of how readily an attack can be reproduced beyond its original study setting. We adopt this criterion instead of impact because impact depends heavily on the target function and deployment context. Below, we detail our replicability criteria and metrics:\looseness-1


\shortsectionEmph{\underline{V}ulnerability \underline{R}estrictiveness  (\prev):} The vulnerability underlying the attack could be \textit{standard (\standard)}: a part of CAN specifications and every system has it if there are no security measures; \textit{widespread (\widespread)}: not part of the standard but present on many CAN systems; \textit{restricted (\specific)}: local only to a certain system, ECU, or higher-layer protocol implementation.\looseness-1

\shortsectionEmph{\underline{D}i\underline{f}ficulty (\diff):} An exploit is \textit{easy (\Circle)} if it does not require pre-existing knowledge or continuous high-overhead computation. It is \textit{moderate (\LEFTcircle)} if it requires pre-existing knowledge, or continuous high-overhead computation. It is \textit{difficult (\CIRCLE)} if it requires compromising more than one ECU, obtaining extensive physical access, or builds upon another {restricted} vulnerability or {difficult} attack.\looseness-1

\shortsectionEmph{Weakest \underline{A}ttacker \underline{M}odel (\abl):} The weakest archetypal attacker model (Sec. \ref{subsec:Models}) required to launch the attack.

\shortsectionBf{Highly Replicable Attack.}  We consider an attack highly replicable if it exploits a {standard} vulnerability (\prev), has {easy} to {moderate} difficulty (\diff), and is remotely launchable (i.e., its weakest attacker model (\abl) is a {bit-banger} or a {by-the-rules} attacker).\looseness-1

\subsection{Confidentiality and Privacy Attacks}

\label{subsec:AttackC}

\mysub{Target: Higher Layer}
\label{subsubsec:AttackCH}
\setcounter{paranum}{0}

  

\confsub{Communication Surveillance} \label{subsubsec:reveng} On a broadcast bus, any attacker could surveil bus messages (\prev: {\standard}). However, the attacker needs format knowledge or reverse engineering to understand their meaning as most vehicle manufacturers keep formats proprietary (\diff: {\LEFTcircle}). While a full understanding of the proprietary formats often requires full physical access using special equipment \cite{yu2022towards}, research shows that an adequate understanding only requires sniffing the entire bus messages or having a trace-dump and then observing correlations, message priorities, fuzzing, probing \cite{markovitz2017field,pese2019librecan, marchetti2019read, lukasiewycz2016security,frassinelli2020know}, or using third-party debuggers and phone companion apps \cite{koscher2010experimental, wen2020automated}. Public datasets of reverse-engineered message formats of many vehicle models are available \cite{opendbc}. Attackers could obtain various private system and personal information from correlating transmitted sensor data, including driver location and vehicle identification, driving patterns profiling, and trip information \cite{wen2020plug,frassinelli2020know, enev2016automobile, fugiglando2017characterizing}. In some systems that reuse the same secret keys for ECU diagnostic authentication (Sec. \ref{sec:background}), attackers could sniff and replay the keys to bypass authentication.\looseness-1

\confsub{Memory Leakage}\label{subsubsec:leakage} {By-the-rule}s attackers can exploit vulnerable higher-layer protocol implementations (\prev: {\specific}) to reveal ECUs' memory contents. They could issue diagnostic commands (Sec. \ref{sec:background}) to access an ECU's sensitive memory, extract keys, or even sometimes dump the entire firmware \cite{koscher2010experimental, milburn2018there, van2021automotive}. However, this requires them to successfully bypass the ECU's diagnostic authentication first (another attack explained in \Ic). Alternatively, Chatterjee et al.~\cite{chatterjee2021exploiting} showed that attackers could reveal memory contents of a victim sender-ECU that uses higher-layer protocols allowing packet fragmentation (e.g., J1939, Sec. \ref{sec:background}). In such multi-frame sessions, the sender first announces the intended transfer length, and the receiver responds with flow-control information that tells the sender what to transmit next. An attacker on the receiving ECU can manipulate the response message, such as requesting more frames than the announced transfer length. In vulnerable implementations, this may cause the victim to transmit additional unintended data beyond the message boundary.\looseness-1

\mysub{Target: Controller Layer}
\label{subsubsec:AttackCD}
\setcounter{paranum}{2}

\confsub{Network Reconnaissance}\label{subsubsec:errorRecon}
The source ECUs of messages are proprietary information of the manufacturer, but {by-the-rules} attackers could identify them using various techniques. \MR{8}One technique exploits loopholes in CAN's error handling mechanism (\prev: {\standard}) \cite{serag2021exposing}. The attacker first pushes an ECU to the error-passive state using the error-state manipulation technique (\Ad), then observes bus messages that display an error-passive behavior. The same technique could also be used as a defense to identify the source of malicious messages \cite{shin2023ridas}. Alternatively, reconnaissance can be performed passively using a technique that was initially published as a defense \cite{cho2016fingerprinting, kulandaivel2019canvas}. The attacker continuously monitors and profiles the cumulative clock skews (small drift in cyclical transmission time) of periodic messages. Messages with the same transmitters will have similar clock skews. However, with generic ECU equipment, accurate and continuous monitoring usually requires high overhead. Moreover, this technique can be effectively evaded if ECUs carefully control their transmission time to mimic other nodes \cite{sagong2018cloaking, ying2019shape}.

\mysub{Target: Transwire Layer}
\setcounter{paranum}{3}

\confsub{Hardware and Topology Profiling}
\label{subsubsec:topology} Due to hardware manufacturing differences, different ECUs' messages often exhibit distinct voltage/time characteristics. Accordingly, {down-to-earther}s could infer information about an ECU's hardware by profiling its voltage characteristics using an oscilloscope or analog-to-digital converter (ADC) \cite{murvay2014source, choi2018identifying,  kneib2020easi, kneib2018scission, foruhandeh2019simple, cho2017viden}, or by measuring the time characteristics of rising and falling edges in a frame using a time-to-digital converter (TDC) \cite{zhou2019btmonitor,ohira2020divider,schell2022asymmetric, ohira2021pli}. Moreover, physical layer features are influenced by ECU locations and bus topology. This allows attackers to identify ECU locations or bus topology changes (e.g., ECU addition or removal) using a voltage sampler \cite{levy2021can, murvay2020tidal, groza2021can}, or by connecting wiring from both ends of the bus to a TDC and monitoring the difference in signal arrival time at the two ends \cite{roeschlin2023edgetdc,SPARTA}.

\confsub{Electromagnetic Emissions}
\label{subsubsec:electrosnif}
{Down-to-earther}s can intercept electromagnetic waves of the CAN bus without being physically connected. From up to 1 meter away using a probe and RF analyzer, researchers showed that CAN electromagnetic emissions are interceptable with enough clarity allowing full reconstruction of frames and opening another door to communication surveillance (\Ca) attacks \cite{teo2021forensic}.\as{ [\abl: $\CIRCLE$; $\mathcal{Q}$: $\CIRCLE$; $\mathcal{X}$: $\LEFTcircle$; \prev: $\CIRCLE$; \diff: $\LEFTcircle$]} 

\begin{tcolorbox}[
 colback=blue!5,
  colframe=black!40,
  fonttitle=\bfseries,
  boxsep=0pt,
  left=2pt,
  right=2pt,
  top=2pt,
  bottom=2pt,
  before skip=4pt,
  after skip=4pt,
]
\shortsectionBf{Confidentiality Takeaway:} Confidentiality is least targeted in pure attack papers (17/50), but most targeted when dual-use defense papers are included (33/67). This shift comes from attacker-identification defenses, whose fingerprinting/profiling methods could be exploited by attackers. It also has the highest share of standard-rooted attacks: 4 of 5 categories.

\end{tcolorbox}

\subsection{Integrity Attacks}
\label{subsec:AttackI}

\mysub{Target: Higher Layer}
\label{subsubsec:AttackIH}
\setcounter{paranum}{0}

\intsub{Fake Data Fabrication}
\label{subsubsec:datafabricate}
A {by-the-rules} attacker controlling an ECU could falsify the data this ECU is supposed to send, including sensor signals, commands, etc. \cite{mazloom2016security, woo2014practical,wen2020plug,miller2015remote, miller2014survey, checkoway2011comprehensive, nie2017free, nie2018ota}. However, to achieve a controlled and intended impact, the attacker must know the format of the data/commands that need to be fabricated. This usually requires reverse engineering (\diff: $\LEFTcircle$). Researchers showed that with careful fabrication, they can control vehicle functions such as the radio, engine, brakes, steering, acceleration, etc. \cite{koscher2010experimental}, or manipulate higher-layer protocols (e.g., UDS, Sec. \ref{sec:background}) \cite{evenchick2015introduction, evenchick2018cantact}.\looseness-1

\intsub{Memory and Firmware Manipulation}\label{subsubsec:memoryManip}
{By-the-rules} attackers could modify a victim ECU's memory to tamper with its data or manipulate the firmware's control flow. This can be achieved by exploiting diagnostic services that allow for memory and firmware modification (e.g., UDS RequestDownload or WriteBlock services, Sec. \ref{sec:background}). However, this is only possible if the ECU does not implement diagnostic authentication (\prev: \specific) \cite{koscher2010experimental}, or its diagnostic authentication can be bypassed as explained in \Ic. An alternative approach is to cause buffer overflow in ECUs with vulnerable higher-layer protocols that implement multi-frame sessions (e.g., J1939, Sec. \ref{sec:background}). The attacker may send more packets than the negotiated number at the beginning of the session ~\cite{chatterjee2021exploiting} or reset the number of frames expected by the receiver to a smaller number \cite{mukherjee2016practical}, so the receiver writes past the boundaries allocated for the session.

\intsub{Bypassing Diagnostic Authentication}\label{subsubsec:bypassUDSAuth}
Researchers investigated ways to bypass diagnostic authentication (Sec. \ref{sec:background}). For {by-the-rule}s attackers, methods including replaying authentication messages obtained from off-the-shelf diagnostic tools \cite{miller2013adventures} and brute-forcing UDS passwords \cite{van2021automotive} could work on some ECUs (\prev: \specific). For {down-to-earther}s, more intrusive methods such as applying a voltage glitch to the ECU's processor so it skips authentication instructions \cite{milburn2018there} were successful for some ECUs. Lauser et al. \cite{lauser2023authenticated} proposed to bypass authentication by hijacking a session post authentication: an OBD-II adapter connecting an external legitimate device to the ECU may act initially as a man-in-the-middle attacker, then hijack the session.

\mysub{Target: Controller Layer}
\label{subsubsec:AttackID}
\setcounter{paranum}{3}

\intsub{Impersonation and Replay} \label{subsubsec:impersonation} {By-the-rules} attackers could impersonate another ECU by sending or replaying messages carrying an identifier of the impersonated victim ECU \cite{miller2013adventures, mazloom2016security, woo2014practical, wen2020plug, miller2015remote, miller2014survey, checkoway2011comprehensive, nie2017free, nie2018ota}. Since CAN does not have an explicit source ID field, the identifier is usually the CAN ID as each CAN ID is typically only transmitted by one ECU. In higher-layer protocols with an explicit source ID field, they could forge the source ID (e.g., J1939, Sec. \ref{sec:background})~\cite{ chatterjee2021exploiting,mukherjee2016practical,burakova2016truck}. {By-the-rule}s attackers may be able to perform such impersonation while successfully evading detection by some types of Intrusion Detection Systems (IDS) \cite{cho2016fingerprinting, sagong2018cloaking, bhatia2021evading}.

\as{ [\abl: $\Circle$; $\mathcal{Q}$: $\Circle$; $\mathcal{X}$: $\Circle$; \prev: $\CIRCLE$; \diff: $\Circle$]} 

\intsub{Frame Hijacking}
\label{subsubsec:frameHijack} If an attacker injects a dominant bit when an error-passive victim is transmitting a recessive bit, the victim's error frames will be invisible to other ECUs. This allows the attacker to finish transmitting the frame, effectively overwriting a part of the victim's message \cite{Hijack}. This could achieve various purposes (e.g., if a secret authentication token is transmitted at the beginning of the message, the attacker can change the message's content while keeping the token intact). This attack can be launched by {bit-banger}s or {by-the-rules} attackers invoking {simultaneous transmission} (Sec. \ref{subsubsec:BRules}).

\intsub{Double Receive}
\label{subsubsec:dblrcv} Per the standard, CAN receivers validate a frame on receiving the second-to-last EOF bit without errors. Transmitters consider it valid if the last bit contains no errors. A {bit-banger} could inject an error in the last EOF bit to exploit this standard mismatch (\prev: \standard) \cite{Hijack}. This makes the transmitter believe the message was not received successfully although it was. In most cases, the transmitter retransmits the message. This could achieve various purposes (e.g., cause a desynchronization when nodes need to agree on the counter values in messages).

\intsub{Unorthodox Frames} 
\label{subsubsec:unOrth} Certain fields in a CAN frame are not completely free to set. For example, the Data Length Code (DLC) should match the actual message length. However, when they are not set correctly, the standard does not specify how they should be detected with the error handling mechanism or provide clear guidelines on how receivers should react (\prev: \standard). {Bit-banger}s could exploit these under-defined parts of the protocol to cause or explore undefined receiver behavior \cite{tang2024eracan}. 
\as{ [\abl: $\LEFTcircle$; $\mathcal{Q}$: $\Circle$; $\mathcal{X}$: $\Circle$; \prev: $\CIRCLE$; \diff: $\Circle$]} 

\intsub{Poly-Semantic Frames}
\label{subsubsec:polysemantic} CAN controllers interpret frames by reading voltage samples at periodic sample points. Controllers' sample points are not necessarily the same (\prev: \widespread). A {bit-banger} that knows at least two ECUs' sample points (\diff: $\LEFTcircle$) could fabricate frames with voltage transitions that fall between the ECUs' sample points so different receivers interpret the frames' content differently \cite{Janus,yue2021cancloak}. However, to make all interpretations of their frames accepted by receivers, they must craft the CRC field so that it is valid for all interpretations. Similarly, CANflict \cite{de2022canflict} showed that many commercial ECUs allow disconnecting the CAN controller from the CAN transceiver and connecting it instead to other peripherals including the SPI, UART, or I2C controllers. {Bit bangers} can transmit messages whose physical form is valid under different communication protocols (e.g., CAN and SPI), but with different meanings. This could be used to achieve various purposes, such as exchanging secret messages that may contain a certain payload when interpreted as a CAN frame, but a different payload when interpreted as another protocol. 

\intsub{Frame Tampering}
\label{subsubsec:frmtmpr}
Attackers may tamper with frames in transit to cause receivers to see a different message from the one sent by the legitimate transmitter. Due to the parasitic reactance of the bus (\prev: \standard), Mohammed et al. \cite{mohammed2022physical} discovered that 8 {bit-banger}s (\diff: $\CIRCLE$) inducing transient signals near a receiver may flip bit values, including 0$\rightarrow$1, on the receiver's side only. The transmitter will not raise a bit error, as it does not detect the flip. The flips should encompass the CRC to accord with the frame, constituting a form of MITM.\as{ [\abl: $\LEFTcircle$; $\mathcal{Q}$: $\Circle$; $\mathcal{X}$: $\Circle$; \prev: $\CIRCLE$; \diff: $\CIRCLE$]} Tang et al. \cite{tang2024eracan} proposed a method requiring only 1 ECU which relies on sample point discrepancy (\prev: \widespread, \diff: $\LEFTcircle$). The attacker injects pulses between the receiver and transmitter sample points to make the receiver read a different message while not triggering the transmitter's error detection. While 1$\rightarrow$0 flips are easy to achieve as CAN physical layer permits it, Rogers et al. showed that one {down-to-earther} may perform  0$\rightarrow$1 flips, allowing it to make active error frames disappear \cite{rogers2022detecting}.

\mysub{Target: Transwire Layer}
\label{subsubsec:AttackIP}
\setcounter{paranum}{9}

\intsub{Physical-Fingerprint Distortion}
\label{subsubsec:volCorr}
{By-the-rule}s attackers could change the physical characteristics of the messages they transmit over the wire to achieve various purposes. They may evade IDSs that use these physical characteristics to detect attacks (physical IDS, explained in \DIi) such as impersonation (\Id). For example, by carefully controlling their transmission time (\diff: $\LEFTcircle$) they may evade IDSs that use clock skews of periodic messages \cite{sagong2018cloaking, ying2019shape}.\as{ [\abl: $\Circle$; $\mathcal{Q}$: $\Circle$; $\mathcal{X}$: $\Circle$; \prev: $\Circle$; \diff: $\LEFTcircle$]} \MR{8}Two {by-the-rules}-controlled ECUs (\diff: $\CIRCLE$) invoking simultaneous transmission (Sec. \ref{subsubsec:simultaneous}) \cite{bhatia2021evading} or a single {bit-banger} (\diff: $\Circle$) \cite{tang2024eracan} can distort an ECU's voltage signature as the resulting signal will be the superposition of the two transmitters' signals. The attacker first does that in the training or the online update period of the IDS to poison its training set to learn the superposition voltage as the victim's signature.

\begin{tcolorbox}[
 colback=blue!5,
  colframe=black!40,
  fonttitle=\bfseries,
  boxsep=0pt,
  left=2pt,
  right=2pt,
  top=2pt,
  bottom=2pt,
  before skip=4pt,
  after skip=4pt
]
\shortsectionBf{Integrity Takeaway:} Most targeted, most diverse, and most replicable. Integrity is the most targeted CIA element, appearing in 28 of 50 papers, and has the highest attack diversity, with 10 categories. It also has the highest number and proportion of highly replicable attacks: 6 of 10.
\end{tcolorbox}

\subsection{Availability Attacks}
\label{subsec:AttackA}

\mysub{Target: Higher Layer}
\setcounter{paranum}{0}

\avsub{ECU Functionality Disabling}\label{subsubsec:ECUDisable}
{By-the-rule}s attackers may disable many ECUs directly by sending higher-layer commands (e.g., UDS, Sec. \ref{sec:background}) to the engine \cite{lee2019enhanced}, brakes, power steering, ignition, lights \cite{koscher2010experimental}, wipers, wireless door locking \cite{wen2020plug}, and other vehicle systems \cite{miller2015remote,nie2017free, miller2013adventures,burakova2016truck}. Attackers can know which diagnostic commands to use by reverse engineering off-the-shelf diagnostic tools that are made for the specific vehicle make and model. Alternatively, they could computationally overwhelm a victim ECU to make it unable to perform its main functionality (e.g., by leading it to miss real-time deadlines). Mukherjee et al.~\cite{mukherjee2016practical} showed that attackers can achieve this exploiting higher-layer protocols that support request-response dynamics (e.g., J1939, Sec. \ref{sec:background}. \prev: \widespread) by simply sending a high number of requests to a victim (\diff: \Circle).

\avsub{Sybil Exhaustion} 
\label{subsubsec:j1939ConExhaust}
Protocols supporting sessions, or that allow ECUs to claim higher-layer addresses (e.g., J1939, Sec. \ref{sec:background}. \prev: \widespread) may be vulnerable to starvation attacks. Attackers can claim the same address as one or more victims to prevent them from getting the desired address \cite{kumar2024security}. Similarly, establishing connections from fake sources with a victim could prevent it from establishing new legitimate connections~\cite{chatterjee2021exploiting}.

\begin{tcolorbox}[
 colback=blue!5,
  colframe=black!40,
  fonttitle=\bfseries,
  boxsep=0pt,
  left=2pt,
  right=2pt,
  top=2pt,
  bottom=2pt,
  before skip=4pt,
  after skip=4pt
]
\shortsectionBf{Higher Layer Takeaway:} Heavily studied but less replicable. Higher-layer attacks appear in 29 of 50 attack papers, second only to controller-layer attacks, yet have a low replication likelihood. They rely least on standard-rooted vulnerabilities and most on restricted ones. Only 2/7 are highly replicable.
\end{tcolorbox}

\mysub{Target: Controller Layer}
\label{subsubsec:AttackAD}
\setcounter{paranum}{2}

\avsub{Error-Injection}
\label{subsubsec:ErrorInj}
Attackers could cause errors in messages transmitted by a victim ECU by inducing controlled collisions to flip a bit from 1 to 0 anywhere past arbitration. This results in an error frame whose direct impacts are destroying the frame and increasing the victim's error counters (Sec. \ref{subsubsec:errors}, \prev: \standard). To do so, {by-the-rule}s attackers could use the simultaneous transmission technique, explained in the by-the-rules model in Sec. \ref{subsubsec:simultaneous}. {Bit-bangers} could flip the bit directly. Error injection could be used for testing \cite{CANstress}, to destroy frames, or as a stepping stone for other attacks (e.g., \Ad).

\avsub{Error State Manipulation}
\label{subsubsec:BusOff}
 Attackers could perform repeated error injection (\Ac) to change a victim's error counter and push it to an error state of their choosing (\prev: \standard). They can push a victim to the error-passive state to prepare for subsequent attacks (e.g., \Cc, \Ie, and \Ij), or to the bus-off state to achieve persistent DoS of the victim. Error state manipulation was first proposed theoretically \cite{wolf2004security}. Later, Cho et al. achieved it with simultaneous transmission (Sec. \ref{subsubsec:simultaneous}) by {by-the-rule}s attackers \cite{cho2016error}. The technique makes an initial collision with one of the victim's messages, which results in consecutive and repeated collisions with all of the victim's retransmission attempts until it becomes error-passive. Then, the attacker launches 19+ rounds of collisions each with a newly transmitted message by the victim until bus-off. \MR{8}Palanca et al. showed that {bit-banger}s could reduce this to a single round, by continuously attacking the victim's retransmissions until bus-off \cite{palanca2017stealth}. Later, CANOX showed that even {by-the-rule}s attackers could do it in one round \cite{serag2021exposing, seragattacks,SeragAlsharif}. However, previous variations rely on attacking message retransmissions and could be slowed down by disabling them (see \DAb). Recently, researchers proposed a single-round non-retransmission-based technique for {bit-banger}s \cite{froschle2017analyzing, serag2023zbcan} called in-frame bus-off.

\avsub{Bus-Access Denial} 
\label{subsubsec:flooding}
{By-the-rule}s attackers could deny other ECUs access to the bus \cite{wolf2004security,koscher2010experimental, miller2013adventures} by flooding it with high-priority messages. An attacker may flood the bus with the highest-priority message to prevent all ECUs from accessing the bus, with a specific-priority message to selectively deny all messages with a lower priority. On the physical level, dominant voltage trumps recessive voltage. As a result, {Bit-banger}s could additionally: 1) inject a continuous dominant voltage to prevent any transmission \cite{murvay2017attacks, koscher2010experimental}, 2) inject a dominant bit after EOF to cause an overload frame (Sec. \ref{sec:background}) then continuously after every overload delimiter to prevent transmission without increasing error counters, or 3) deny arbitration for specific messages only by flipping a 1 to a 0 in the IDs ~\cite{murvay2017attacks,de2022canflict,humayed2020cansentry}.\looseness-1

\avsub{Synchronization Disruption}
\label{subsubsec:synchDis}
{Bit-banger}s could disrupt synchronization among different ECUs on the bus to cause various communication errors.
Since CAN controllers use dynamic edge-based synchronization, attackers could exploit this (\prev: \standard) by injecting accurately-timed pulses within bits during frame transmission~\cite{murvay2017attacks,de2022canflict, tang2024eracan}. This leads ECUs to resynchronize based on the edges of those pulses and misread the beginnings and ends of different bits.


\begin{tcolorbox}[
 colback=blue!5,
  colframe=black!40,
  fonttitle=\bfseries,
  boxsep=0pt,
  left=2pt,
  right=2pt,
  top=2pt,
  bottom=2pt,
  before skip=4pt,
  after skip=4pt,
]
\shortsectionBf{Controller Layer Takeaway:} Most targeted, most diverse, and most replicable. The controller layer dominates attack interest, with 32 of 50 attack papers and 11 attack categories. Its attacks mostly exploit standard-rooted vulnerabilities, none rely on restricted ones, and most are easy to launch. Thus, 9 of 11 categories are highly replicable.

\end{tcolorbox}

\mysub{Target: Transwire Layer}
\setcounter{paranum}{6}

\avsub{Signal Attenuation}\label{subsubsec:sigAttenuate} Mohammed et al. \cite{mohammed2022physical} discovered that 8 {bit-bangers} (\diff: $\CIRCLE$) inducing transient signals near a receiver can attenuate the voltage signal between a transmitter and a receiver, possibly causing the message to be invisible to selected ECUs. This technique becomes more effective as the distance between the transmitter and the receiver grows (\prev: \standard).\looseness-1

\begin{tcolorbox}[
 colback=blue!5,
  colframe=black!40,
  fonttitle=\bfseries,
  boxsep=0pt,
  left=2pt,
  right=2pt,
  top=2pt,
  bottom=2pt,
  before skip=4pt,
  after skip=4pt,
]
\shortsectionBf{Transwire Layer Takeaway:} Least targeted and least diverse. The transwire layer appears in only 6 of 50 attack papers and spans 4 attack categories. All exploit standard-rooted vulnerabilities, but only one is highly replicable due to difficult attack steps or stronger attacker requirements.
\end{tcolorbox}

\begin{tcolorbox}[
 colback=blue!5,
  colframe=black!40,
  fonttitle=\bfseries,
  boxsep=0pt,
  left=2pt,
  right=2pt,
  top=2pt,
  bottom=2pt,
  before skip=4pt,
  after skip=4pt,
]
\shortsectionBf{Availability Takeaway:} Frequently targeted and broadly replicable. Availability ranks second in attack papers (22 of 50), attack diversity (7 categories), and highly replicable attacks (4 of 7). All attacks exploit standard-rooted or widespread vulnerabilities; none rely on restricted vulnerabilities.


\end{tcolorbox}






\definecolor{Matrix}{rgb}{0.65,0.301,0.301}
\definecolor{BeautyBush}{rgb}{0.90,0.70,0.60}
\definecolor{Manhattan}{rgb}{0.7600,0.7955,0.8992}
\definecolor{BurntOrange}{rgb}{0.3373, 0.4353, 0.7216}
\definecolor{Bobo}{RGB}{176,145,122}

\definecolor{SecureEfficient}{RGB}{34,139,34}
\definecolor{SecureOnly}{RGB}{140,150,60}
\definecolor{EfficientOnly}{RGB}{255,200,0}
\definecolor{Limited}{RGB}{195,100,210}

\newcolumntype{D}{>{\centering\arraybackslash}m{0.78cm}}

\newcommand{\datk}[1]{%
  \rotatebox[origin=c]{90}{\parbox{2.2cm}{\centering\scriptsize #1}}}

\newcommand{\tblhead}[1]{{\footnotesize\bfseries #1}}
\newcommand{\layerhead}[1]{{\scriptsize\bfseries #1}}

\newlength{\DefenseRowHeight}
\setlength{\DefenseRowHeight}{0.42cm}
\newsavebox{\DefenseNameBox}
\newcommand{\dname}[1]{%
  \parbox[c][\DefenseRowHeight][c]{3.18cm}{%
    \centering\scriptsize
    \sbox{\DefenseNameBox}{\mbox{#1}}%
    \ifdim\wd\DefenseNameBox>3.08cm
      \resizebox{3.08cm}{!}{\usebox{\DefenseNameBox}}%
    \else
      \usebox{\DefenseNameBox}%
    \fi}}

\newlength{\DefenseColumnHeaderHeight}
\setlength{\DefenseColumnHeaderHeight}{0.36cm}
\newcommand{\columncategory}[1]{%
  \parbox[c][\DefenseColumnHeaderHeight][c]{2.3cm}{\centering\tblhead{#1}}}
\newcommand{\columnlayer}[1]{%
  \parbox[c][\DefenseColumnHeaderHeight][c]{0.55cm}{\centering\layerhead{#1}}}
\newcommand{\rowcategory}[3]{%
  \multirow[c]{#1}{0.36cm}{%
    \centering\rotatebox[origin=c]{90}{\tblhead{#3}}}}
\newcommand{\rowlayergroup}[3]{%
  \multirow[c]{#1}{0.36cm}{%
    \centering\rotatebox[origin=c]{90}{\layerhead{#3}}}}
\newcommand{\rowlayer}[1]{%
  \parbox[c][\DefenseRowHeight][c]{0.36cm}{%
    \centering\rotatebox[origin=c]{90}{\layerhead{#1}}}}

\newlength{\DefenseHeaderWidth}
\newlength{\DefenseHeaderHeight}
\newcommand{\defensecorner}{%
  \diagbox[
    width=\DefenseHeaderWidth,
    height=\DefenseHeaderHeight,
    dir=NW,
    innerleftsep=0.20cm,
    innerrightsep=0.20cm,
    outerleftsep=0pt,
    outerrightsep=0pt,
    linewidth=\arrayrulewidth,
    font=\footnotesize\bfseries
  ]{\raisebox{0.16cm}{Defenses}}{\raisebox{-0.16cm}{Attacks}}}

\expandafter\def\csname assessment@green@SecureEfficient\endcsname{}
\expandafter\def\csname assessment@ideal@insurance@P\endcsname{}
\expandafter\def\csname assessment@ideal@coverage@Wide\endcsname{}
\expandafter\def\csname assessment@ideal@validity@Evaluated\endcsname{}
\expandafter\def\csname assessment@ideal@validity@Analyzed\endcsname{}
\expandafter\def\csname assessment@ideal@level@High\endcsname{}
\expandafter\def\csname assessment@ideal@level@Moderate\endcsname{}

\newcommand{\MissingCriterionSymbol}[1]{%
  \raisebox{0.05ex}{\scalebox{0.62}{\textsf{\bfseries #1}}}}
\newcommand{\AssessmentMissingSymbol}[3]{%
  \ifcsname assessment@ideal@#1@#2\endcsname
  \else
    \MissingCriterionSymbol{#3}%
  \fi}
\newcommand{\AssessmentMissingSymbols}[6]{%
  \AssessmentMissingSymbol{insurance}{#1}{P}%
  \AssessmentMissingSymbol{coverage}{#2}{C}%
  \AssessmentMissingSymbol{validity}{#3}{V}%
  \AssessmentMissingSymbol{level}{#4}{E}%
  \AssessmentMissingSymbol{level}{#5}{B}%
  \AssessmentMissingSymbol{level}{#6}{I}}

\newcommand{\acell}[7]{%
  \cellcolor{#1}%
  \parbox[c][\DefenseRowHeight][c]{\linewidth}{%
    \centering
    \ifcsname assessment@green@#1\endcsname
      \mbox{}%
    \else
      \mbox{\AssessmentMissingSymbols{#2}{#3}{#4}{#5}{#6}{#7}}%
    \fi}}

\newcommand{\acellNoPreventionMarker}[7]{%
  \acell{#1}{P}{#3}{#4}{#5}{#6}{#7}}

\newcommand{\dashcell}[1]{%
  \cellcolor{#1}%
  \parbox[c][\DefenseRowHeight][c]{\linewidth}{%
    \centering\mbox{\ensuremath{-}}}}

\newcommand{\VerdictSample}[1]{%
  \tikz[baseline=-0.45ex]{\draw[fill=#1,line width=0.3pt]
    (0,0) rectangle (0.20cm,0.12cm);}}
\newcommand{\LegendLine}[1]{%
  \parbox[c][\DefenseRowHeight][c]{20.8cm}{%
    \centering\scriptsize\strut #1\strut}}
\newcommand{\ColorLegendEntry}[2]{%
  \mbox{\raisebox{-0.12ex}{\VerdictSample{#1}}\hspace{0.25em}\strut #2}}
\newcommand{\SymbolLegendEntry}[2]{%
  \mbox{\MissingCriterionSymbol{#1}\hspace{0.20em}\strut #2}}

\begin{table*}[!t]
\centering
\caption{CAN Defense Taxonomy and Effectiveness Assessment per Attack Category\\%
  \textbf{HL}: Higher Layer,\quad \textbf{CT}: Controller Layer,\quad \textbf{TW}: Transwire Layer}
\label{tbl:defense_taxonomy}
\label{tbl:bigtable}
\resizebox{\textwidth}{!}{%
\setlength{\tabcolsep}{0.5pt}
\renewcommand{\arraystretch}{1.00}
\setlength{\DefenseHeaderWidth}{%
  \dimexpr4.02cm+6\tabcolsep+2\arrayrulewidth\relax}
\setlength{\DefenseHeaderHeight}{2.92cm}
\begin{tabular}{%
  |>{\centering\arraybackslash}m{0.36cm}
  |>{\centering\arraybackslash}m{0.36cm}
  |>{\centering\arraybackslash}m{3.30cm}
  |D|D|D|D|D
  |D|D|D|D|D|D|D|D|D|D
  |D|D|D|D|D|D|D|}
\hhline{|-|-|-|-|-|-|-|-|-|-|-|-|-|-|-|-|-|-|-|-|-|-|-|-|-|}
\multicolumn{3}{|@{}c@{}|}{\multirow{3}{*}{\defensecorner}} &
  \multicolumn{5}{c|}{\columncategory{Confidentiality}} &
  \multicolumn{10}{c|}{\columncategory{Integrity}} &
  \multicolumn{7}{c|}{\columncategory{Availability}} \\
\hhline{|~~~|-|-|-|-|-|-|-|-|-|-|-|-|-|-|-|-|-|-|-|-|-|-|}
\multicolumn{3}{|c|}{} &
  \multicolumn{2}{c|}{\columnlayer{HL}}     &
  \multicolumn{1}{c|}{\columnlayer{CT}}     &
  \multicolumn{2}{c|}{\columnlayer{TW}}     &
  \multicolumn{3}{c|}{\columnlayer{HL}}     &
  \multicolumn{6}{c|}{\columnlayer{CT}}     &
  \multicolumn{1}{c|}{\columnlayer{TW}}     &
  \multicolumn{2}{c|}{\columnlayer{HL}}     &
  \multicolumn{4}{c|}{\columnlayer{CT}}     &
  \multicolumn{1}{c|}{\columnlayer{TW}}     \\
\hhline{|~~~|-|-|-|-|-|-|-|-|-|-|-|-|-|-|-|-|-|-|-|-|-|-|}
\multicolumn{3}{|c|}{} &
  \datk{\cellcolor{Matrix}\textcolor{white}{\Ca.\ Comm.\ Surveillance}}     &
  \datk{\Cb. Memory Leakage}          &
  \datk{\cellcolor{BurntOrange}\textcolor{white}{\Cc. Network Reconnaissance}}    &
  \datk{\Cd. HW/Topology Profiling}    &
  \datk{\Ce. EM Emissions}            &
  \datk{\cellcolor{Matrix}\textcolor{white}{\Ia. Fake Data Fabrication}}   &
  \datk{\Ib. Mem./FW Manipulation}    &
  \datk{\Ic. Bypass Diag.\ Authentication}     &
  \datk{\cellcolor{Bobo}\Id. Impersonation \& Replay} &
  \datk{\cellcolor{Bobo}\Ie. Frame Hijacking}         &
  \datk{\cellcolor{BurntOrange}\textcolor{white}{\IF. Double Receive}}          &
  \datk{\cellcolor{Manhattan}\Ig. Unorthodox Frames}       &
  \datk{\Ih. Poly-Semantic Frames}    &
  \datk{\Ii. Frame Tampering}         &
  \datk{\cellcolor{Manhattan}\Ij. Physical FP Distortion}   &
  \datk{\Aa. ECU Func.\ Disabling}   &
  \datk{\Ab. Sybil Exhaustion}        &
  \datk{\cellcolor{BurntOrange}\textcolor{white}{\Ac. Error Injection}}         &
  \datk{\cellcolor{BurntOrange}\textcolor{white}{\Ad. Error State Manipulation}}      &
  \datk{\cellcolor{BurntOrange}\textcolor{white}{\Ae. Bus Access Denial}}       &
  \datk{\cellcolor{BurntOrange}\textcolor{white}{\Af. Synchronization Disruption}}      &
  \datk{\Ag. Signal Attenuation}      \\
\hhline{|-|-|-|-|-|-|-|-|-|-|-|-|-|-|-|-|-|-|-|-|-|-|-|-|-|}

\rowcategory{3}{-0.04cm}{Conf.}
& \rowlayer{HL}
& \dname{\DCa.\ Encryption \& Obfuscation}
& \acell{Limited}{P}{Restricted}{Evaluated}{Low}{Low}{High}
& ~  & ~  & ~  & ~
& ~ & ~ & ~ & ~ & ~ & ~ & ~ & ~ & ~ & ~
& ~ & ~ & ~ & ~ & ~ & ~ & ~
\\
\hhline{|~|-|-|-|-|-|-|-|-|-|-|-|-|-|-|-|-|-|-|-|-|-|-|-|-|}
& \rowlayer{CT}
& \dname{\DCb.\ ID Anonymization}
& ~
& ~
& \acell{EfficientOnly}{P}{Restricted}{Evaluated}{High}{High}{High}
& ~  & ~
& ~ & ~ & ~
& ~
& \acell{EfficientOnly}{P}{Restricted}{Evaluated}{High}{High}{High}
& ~  & ~  & ~  & ~
& \acell{EfficientOnly}{P}{Restricted}{Evaluated}{High}{High}{High}
& ~  & ~
& \acell{EfficientOnly}{P}{Restricted}{Evaluated}{High}{High}{High}
& \acell{EfficientOnly}{P}{Restricted}{Evaluated}{High}{High}{High}
& ~  & ~  & ~
\\
\hhline{|~|-|-|-|-|-|-|-|-|-|-|-|-|-|-|-|-|-|-|-|-|-|-|-|-|}
& \rowlayer{TW}
& \dname{\DCc.\ Shielding \& Slope Control}
& ~  & ~  & ~  & ~
& \cellcolor{Limited}\parbox[c][\DefenseRowHeight][c]{\linewidth}{%
    \centering\mbox{\MissingCriterionSymbol{V}\MissingCriterionSymbol{I}}}
& ~ & ~ & ~ & ~ & ~ & ~ & ~ & ~ & ~ & ~
& ~ & ~ & ~ & ~ & ~ & ~ & ~
\\
\hhline{|-|-|-|-|-|-|-|-|-|-|-|-|-|-|-|-|-|-|-|-|-|-|-|-|-|}

\rowcategory{10}{1.79cm}{Integrity} & \rowlayergroup{3}{0.44cm}{HL}
& \dname{\DIb.\ Content Inspection}
& ~  & ~  & ~  & ~  & ~
& \acell{EfficientOnly}{D}{Restricted}{Evaluated}{High}{High}{High}
& \acell{EfficientOnly}{DStar}{Wide}{Evaluated}{High}{High}{High}
& ~
& ~
& ~  & ~  & ~  & ~  & ~  & ~
& \acell{EfficientOnly}{D}{Wide}{Analyzed}{High}{High}{High}
& \acell{EfficientOnly}{D}{Wide}{Analyzed}{High}{High}{High}
& ~ & ~ & ~ & ~ & ~
\\
\hhline{|~|~|-|-|-|-|-|-|-|-|-|-|-|-|-|-|-|-|-|-|-|-|-|-|-|}
& &
  \dname{\DIc.\ Patching \& Implementation}
& ~
& \acell{SecureEfficient}{P}{Wide}{Evaluated}{High}{High}{High}
& ~  & ~  & ~
& ~
& \acell{SecureEfficient}{P}{Wide}{Evaluated}{High}{High}{High}
& \acell{SecureEfficient}{P}{Wide}{Evaluated}{High}{High}{High}
& ~  & ~  & ~  & ~  & ~  & ~  & ~
& \acell{SecureEfficient}{P}{Wide}{Analyzed}{High}{High}{High}
& \acell{SecureEfficient}{P}{Wide}{Analyzed}{High}{High}{High}
& ~ & ~ & ~ & ~ & ~
\\
\hhline{|~|~|-|-|-|-|-|-|-|-|-|-|-|-|-|-|-|-|-|-|-|-|-|-|-|}
& &
  \dname{\DId.\ ECU FW Hardening}
& ~
& \acell{EfficientOnly}{P}{Restricted}{Analyzed}{Moderate}{High}{Moderate}
& ~  & ~  & ~
& ~
& \acell{EfficientOnly}{P}{Restricted}{Analyzed}{Moderate}{High}{Moderate}
& ~  & ~  & ~  & ~  & ~  & ~  & ~  & ~
& ~ & ~ & ~ & ~ & ~ & ~ & ~
\\
\hhline{|~|-|-|-|-|-|-|-|-|-|-|-|-|-|-|-|-|-|-|-|-|-|-|-|-|}
& \rowlayergroup{4}{0.63cm}{CT}
& \dname{\DIa.\ Message Authentication}
& ~ & ~ & ~ & ~ & ~
& ~ & ~
& \acell{SecureEfficient}{P}{Wide}{Evaluated}{High}{High}{High}
& \acell{SecureEfficient}{DStar}{Wide}{Evaluated}{Moderate}{Moderate}{High}
& ~  & ~  & ~  & ~  & ~  & ~
& ~ & ~ & ~ & ~ & ~ & ~ & ~
\\
\hhline{|~|~|-|-|-|-|-|-|-|-|-|-|-|-|-|-|-|-|-|-|-|-|-|-|-|}
& &
  \dname{\DIF.\ Victim Resistance}
& ~ & ~ & ~ & ~ & ~
& ~ & ~ & ~
& \acell{SecureEfficient}{P}{Wide}{Evaluated}{High}{High}{Moderate}
& ~  & ~  & ~  & ~  & ~  & ~
& ~ & ~ & ~ & ~ & ~ & ~ & ~
\\
\hhline{|~|~|-|-|-|-|-|-|-|-|-|-|-|-|-|-|-|-|-|-|-|-|-|-|-|}
& &
  \dname{\DIi.\ Physical IDS}
& ~ & ~ & ~ & ~ & ~
& ~ & ~ & ~
& \acell{SecureEfficient}{DStar}{Wide}{Evaluated}{High}{High}{High}
& ~  & ~  & ~  & ~
& \acell{EfficientOnly}{D}{Restricted}{Evaluated}{High}{High}{High}
& ~
& ~  & ~
& \acell{EfficientOnly}{D}{Restricted}{Evaluated}{High}{High}{High}
& \acell{EfficientOnly}{D}{Restricted}{Evaluated}{High}{High}{High}
& ~  & ~
& \acell{EfficientOnly}{D}{Restricted}{Analyzed}{High}{High}{High}
\\
\hhline{|~|~|-|-|-|-|-|-|-|-|-|-|-|-|-|-|-|-|-|-|-|-|-|-|-|}
& &
\dname{\DIg.\ Unify Sample Points}
& ~ & ~ & ~ & ~ & ~
& ~ & ~ & ~ & ~ & ~ & ~ & ~
& \acell{EfficientOnly}{P}{Restricted}{Analyzed}{High}{High}{Moderate}
& \acell{EfficientOnly}{P}{Restricted}{Analyzed}{High}{High}{Moderate}
& ~
& ~ & ~ & ~ & ~ & ~ & ~ & ~
\\
\hhline{|~|-|-|-|-|-|-|-|-|-|-|-|-|-|-|-|-|-|-|-|-|-|-|-|-|}
& \rowlayergroup{3}{0.44cm}{TW}
& \dname{\DIh.\ Time \& IFS Anonymization}
& ~  & ~
& \acell{EfficientOnly}{P}{Restricted}{Evaluated}{Moderate}{Moderate}{High}
& ~  & ~
& ~ & ~ & ~
& \acell{SecureEfficient}{DStar}{Wide}{Evaluated}{Moderate}{Moderate}{High}
& \acell{EfficientOnly}{P}{Restricted}{Evaluated}{Moderate}{Moderate}{High}
& ~  & ~  & ~  & ~
& \acell{EfficientOnly}{P}{Restricted}{Evaluated}{Moderate}{Moderate}{High}
& ~  & ~
& \acell{EfficientOnly}{P}{Restricted}{Evaluated}{Moderate}{Moderate}{High}
& \acell{EfficientOnly}{P}{Restricted}{Evaluated}{Moderate}{Moderate}{High}
& ~  & ~  & ~
\\
\hhline{|~|~|-|-|-|-|-|-|-|-|-|-|-|-|-|-|-|-|-|-|-|-|-|-|-|}
& &
  \dname{\DIj.\ Bit-Banger IDS}
& ~ & ~ & ~ & ~ & ~
& ~ & ~ & ~
& \acell{SecureEfficient}{DStar}{Wide}{Evaluated}{High}{High}{High}
& \acell{SecureEfficient}{DStar}{Wide}{Evaluated}{High}{High}{High}
& \acell{EfficientOnly}{D}{Wide}{Evaluated}{High}{High}{High}
& \acell{SecureEfficient}{DStar}{Wide}{Evaluated}{High}{High}{High}
& \acell{SecureEfficient}{DStar}{Wide}{Evaluated}{High}{High}{High}
& \acellNoPreventionMarker{EfficientOnly}{DStar}{Restricted}{Evaluated}{High}{High}{High}
& \acell{SecureEfficient}{DStar}{Wide}{Evaluated}{High}{High}{High}
& ~  & ~
& \acell{EfficientOnly}{D}{Wide}{Evaluated}{High}{High}{High}
& \acell{EfficientOnly}{D}{Wide}{Evaluated}{High}{High}{High}
& \acell{EfficientOnly}{D}{Wide}{Evaluated}{High}{High}{High}
& \acell{EfficientOnly}{D}{Wide}{Evaluated}{High}{High}{High}
& ~
\\
\hhline{|~|~|-|-|-|-|-|-|-|-|-|-|-|-|-|-|-|-|-|-|-|-|-|-|-|}
& &
  \dname{\DIk.\ ECU Guardian}
& ~ & ~ & ~ & ~ & ~
& ~
& \acell{SecureOnly}{P}{Wide}{Analyzed}{High}{High}{Low}
& ~
& \acell{SecureOnly}{P}{Wide}{Evaluated}{High}{High}{Low}
& \acell{SecureOnly}{P}{Wide}{Evaluated}{High}{High}{Low}
& \acell{SecureOnly}{P}{Wide}{Evaluated}{High}{High}{Low}
& \acell{SecureOnly}{P}{Wide}{Evaluated}{High}{High}{Low}
& \acell{SecureOnly}{P}{Wide}{Evaluated}{High}{High}{Low}
& \acell{SecureOnly}{P}{Wide}{Evaluated}{High}{High}{Low}
& \acell{SecureOnly}{P}{Wide}{Evaluated}{High}{High}{Low}
& \acell{SecureOnly}{P}{Wide}{Evaluated}{High}{High}{Low}
& \acell{SecureOnly}{P}{Wide}{Evaluated}{High}{High}{Low}
& \acell{SecureOnly}{P}{Wide}{Evaluated}{High}{High}{Low}
& \acell{SecureOnly}{P}{Wide}{Evaluated}{High}{High}{Low}
& & &
\\
\hhline{|-|-|-|-|-|-|-|-|-|-|-|-|-|-|-|-|-|-|-|-|-|-|-|-|-|}

\rowcategory{5}{0.83cm}{Availability} & \rowlayergroup{4}{0.63cm}{CT} &
  \dname{\DAc. Error Handling IDS}
& ~  & ~
& \acell{EfficientOnly}{D}{Wide}{Evaluated}{High}{High}{High}
& ~  & ~
& ~ & ~ & ~
& ~
& \acell{EfficientOnly}{D}{Wide}{Evaluated}{High}{High}{High}
& ~  & ~  & ~  & ~
& \acell{EfficientOnly}{D}{Restricted}{Evaluated}{High}{High}{High}
& ~  & ~  & ~
& \acell{EfficientOnly}{D}{Wide}{Evaluated}{High}{High}{High}
& ~  & ~  & ~
\\
\hhline{|~|~|-|-|-|-|-|-|-|-|-|-|-|-|-|-|-|-|-|-|-|-|-|-|-|}
& &
  \dname{\DAd.\ L2 Node Suspension}
& ~ & ~ & ~ & ~ & ~
& ~ & ~ & ~ & ~ & ~ & ~ & ~ & ~ & ~ & ~
& ~  & ~
& \acell{EfficientOnly}{P}{Restricted}{Evaluated}{High}{High}{High}
& \acell{EfficientOnly}{P}{Restricted}{Evaluated}{High}{High}{High}
& \acell{EfficientOnly}{P}{Restricted}{Evaluated}{High}{High}{High}
& ~  & ~
\\
\hhline{|~|~|-|-|-|-|-|-|-|-|-|-|-|-|-|-|-|-|-|-|-|-|-|-|-|}
& &
\dname{\DAa.\ ID-Priority Dissociation}
& ~ & ~ & ~ & ~ & ~
& ~ & ~ & ~ & ~ & ~ & ~ & ~ & ~ & ~ & ~
& ~  & ~
& \acell{Limited}{P}{Restricted}{Analyzed}{Moderate}{Low}{Low}
& \acell{Limited}{P}{Restricted}{Analyzed}{Moderate}{Low}{Low}
& \acell{Limited}{P}{Restricted}{Analyzed}{Moderate}{Low}{Low}
& ~  & ~
\\
\hhline{|~|~|-|-|-|-|-|-|-|-|-|-|-|-|-|-|-|-|-|-|-|-|-|-|-|}
& &
  \dname{\DAb.\ One-Shot Mode}
& ~  & ~
& \acell{EfficientOnly}{S}{Restricted}{Evaluated}{High}{Moderate}{High}
& ~  & ~
& ~ & ~ & ~
& ~
& \acell{EfficientOnly}{S}{Restricted}{Evaluated}{High}{Moderate}{High}
& ~  & ~  & ~  & ~
& \acell{EfficientOnly}{S}{Restricted}{Evaluated}{High}{Moderate}{High}
& ~  & ~  & ~
& \acell{EfficientOnly}{S}{Restricted}{Evaluated}{High}{Moderate}{High}
& ~  & ~  & ~
\\
\hhline{|~|-|-|-|-|-|-|-|-|-|-|-|-|-|-|-|-|-|-|-|-|-|-|-|-|}
& \rowlayer{TW}
& \dname{\DAe.\ Dynamic Fragmentation}
& ~ & ~ & ~ & ~ & ~
& ~ & ~ & ~
& ~
& \acellNoPreventionMarker{SecureOnly}{Star}{Wide}{Analyzed}{High}{High}{Low}
& \acellNoPreventionMarker{SecureOnly}{Star}{Wide}{Analyzed}{High}{High}{Low}
& \acellNoPreventionMarker{SecureOnly}{Star}{Wide}{Analyzed}{High}{High}{Low}
& \acellNoPreventionMarker{SecureOnly}{Star}{Wide}{Analyzed}{High}{High}{Low}
& \acellNoPreventionMarker{SecureOnly}{Star}{Wide}{Analyzed}{High}{High}{Low}
& \acellNoPreventionMarker{SecureOnly}{Star}{Wide}{Analyzed}{High}{High}{Low}
& ~  & ~
& \acellNoPreventionMarker{SecureOnly}{Star}{Wide}{Evaluated}{High}{High}{Low}
& \acellNoPreventionMarker{SecureOnly}{Star}{Wide}{Evaluated}{High}{High}{Low}
& \acellNoPreventionMarker{SecureOnly}{Star}{Wide}{Evaluated}{High}{High}{Low}
& \acellNoPreventionMarker{SecureOnly}{Star}{Wide}{Evaluated}{High}{High}{Low}
& \acellNoPreventionMarker{SecureOnly}{Star}{Wide}{Analyzed}{High}{High}{Low}
\\
\hhline{|-|-|-|-|-|-|-|-|-|-|-|-|-|-|-|-|-|-|-|-|-|-|-|-|-|}
\multicolumn{25}{|c|}{%
  \LegendLine{Color:\quad
  \ColorLegendEntry{SecureEfficient}{Secure \& Efficient}\quad
  \ColorLegendEntry{SecureOnly}{Secure}\quad
  \ColorLegendEntry{EfficientOnly}{Efficient}\quad
  \ColorLegendEntry{Limited}{Limited}}}\\
\hhline{|-|-|-|-|-|-|-|-|-|-|-|-|-|-|-|-|-|-|-|-|-|-|-|-|-|}
\multicolumn{25}{|c|}{%
  \LegendLine{Lacking Criteria:\hspace{0.7em}
  \SymbolLegendEntry{P}{No Prevention}\hspace{0.7em}
  \SymbolLegendEntry{C}{Restricted Coverage}\hspace{0.7em}
  \SymbolLegendEntry{V}{Limited Validation}\hspace{0.7em}
  \SymbolLegendEntry{E}{Low ECU Efficiency}\hspace{0.7em}
  \SymbolLegendEntry{B}{Low Bus Performance}\hspace{0.7em}
  \SymbolLegendEntry{I}{Low Infrastructure Compatibility}}}\\
\hhline{|-|-|-|-|-|-|-|-|-|-|-|-|-|-|-|-|-|-|-|-|-|-|-|-|-|}
\end{tabular}%
}
\end{table*}

\section{CAN Defenses}
\label{sec:Defenses}

In this section, we taxonomize CAN defenses as described in Sec. \ref{sec:methodology}. For each defense category, we assess its efficacy against relevant attacks and summarize the variants of the same defense approach reported in the literature, along with the challenges they face.  Tbl.~\ref{tbl:defense_taxonomy} summarizes all defenses and juxtaposes them against attacks to reveal defense gaps. The top row lists all attack categories. Highly replicable attacks are highlighted in light brown and light blue for {by-the-rules} attackers and {bit-bangers}, respectively. \MR{2\&9} Each attack $\times$ defense cell reports the defense's assessment against that attack. A green cell marks an effective defense. A non-green cell means a defense can protect against an attack but not effectively. Letters in them list the criteria that the defense lacks, preventing it from being effective.  We provide a spreadsheet at \cite{code} detailing each assessment and its rationale.

\label{subsec:DefenseAssessment}
\shortsectionBf{Defense Assessment Criteria.} We assess defense efficacy per attack. If a defense applies to multiple attacks, its assessment will vary across attacks based on the metrics below:\looseness-1  



\shortsectionEmph{\underline{I}nsu\underline{r}ance (\ins):} A defense insures \textit{prevention} \textit{(P)}, \textit{detection} \textit{(D)}, or \textit{slowing} \textit{(S)} by reducing its speed or effectiveness. \textit{(*)} means the defense insures prevention only when combined with another approach.  



\shortsectionEmph{\underline{C}o\underline{v}erage (\cov):} Assesses how comprehensively a defense addresses different manifestations of an attack, including various attack paths or different attacker models. A defense is \textit{restricted} ($\LEFTcircle$) when it only covers certain manifestations, leaving out known blind spots or attacker models. \textit{Wide} coverage ($\CIRCLE$) means it addresses most or all variations and attackers.\looseness-1 


\shortsectionEmph{\underline{E}CU Operational \underline{E}fficiency (\ecu):} The efficiency is \textit{low} ($\Circle$) when it involves computationally expensive operations for every transmitted and received message. It is \textit{moderate} ($\LEFTcircle$) if such operations are necessary only for some frequent messages or only outgoing ones. It is \textit{high} ($\CIRCLE$) if they are needed only initially or sporadically with infrequent messages.\looseness-1  

\shortsectionEmph{\underline{B}us \underline{P}erformance (\bus):} The performance is \textit{low} ($\Circle$) if it causes dropping or unschedulability (inability to put an upper bound on the maximum message delay), \textit{moderate} ($\LEFTcircle$) if it is noticeable without causing message unschedulability, and \textit{high} ($\CIRCLE$) if its busload or delay is barely noticeable. 

\shortsectionEmph{\underline{I}nfrastructure \underline{C}ompatibility (\infra):} The compatibility is \textit{low} if it requires significant hardware modifications to the bus or each ECU ($\Circle$), \textit{moderate} ($\LEFTcircle$) if it involves some extensive firmware updates or a combination of changes, and \textit{high} ($\CIRCLE$) if it requires limited changes, including adding a monitor node or making small firmware/software updates.

\shortsectionEmph{\underline{V}a\underline{l}idity (\val):} A defense might be \textit{speculated} to work against an attack without being analyzed for CAN ($\Circle$), proposed and \textit{analyzed} for CAN systems but not evaluated ($\LEFTcircle$), or \textit{evaluated} on CAN buses in real or test environments ($\CIRCLE$).\looseness-1

\shortsectionBf{Effective Defense.} We consider a defense effective if it offers at least \textit{prevention} (\ins), \textit{wide} coverage (\cov), \textit{analyzed} validation (\val), and \textit{moderate} ECU operation efficiency (\ecu), bus performance (\bus), and infrastructure compatibility (\infra). 

\begin{tcolorbox}[
  colback=pink!15,
  colframe=pink!60,
fonttitle=\bfseries,
boxsep=0pt,
left=2pt,
right=2pt,
top=2pt,
bottom=2pt,
before skip=4pt,
after skip=4pt,
]
\shortsectionBf{Gaps:} We define a {defense gap} as a highly replicable attack with no effective defense. \MR{10}Tbl. \ref{tbl:defense_taxonomy} highlights {by-the-rules} gaps in dark red, and {bit-banger} gaps in dark blue.
\end{tcolorbox}

\subsection{Confidentiality and Privacy Defenses}
\label{subsec:DefenseC}

\mysub{Protection: Higher Layer}
\setcounter{paranum}{0}

\Dconfsub{Encryption and Obfuscation}\label{subsubsec:encryption}
\label{subsubsec:payloadObf}
Encryption can {prevent} communication surveillance (\Ca) with little infrastructure change (\infra: $\high$). However, all cryptography-based approaches on CAN face 3 challenges: {ECU performance}, {CAN message-length}, and {shared media key management}. When encryption is used extensively, these challenges cause coverage, bus, and ECU efficiency issues. When used sporadically or outside normal operation (e.g., diagnostic sessions), they may not present a hurdle. Since these challenges are common to all cryptographic approaches, they are discussed in detail in Sec. \ref{subsec:defense_rc}. To overcome the length and performance challenges, researchers proposed the following:\looseness-1

\shortsectionEmph{Message Length.} To get over CAN's 64-bit message length, researchers proposed using stream ciphers (e.g., RC4) \cite{lu2019leap}, 64-bit encryption \cite{jukl2016using, bella2019toucan}, truncating 128-bit block ciphers to 64 bits to generate pads and XORing them \cite{woo2014practical}, or sending the payload over two messages \cite{mundhenk2017security,pfeiffer2017implementing}. However, this increases the busload as it doubles each message (\bus: \low) or forces it to adopt the maximum length (\bus: \moderate).\looseness-1

\shortsectionEmph{Performance.} Some techniques attempt to reduce the overhead and busload impact (\ecu: \moderate, \bus: \high) by using non-encryption based obfuscation, such as Fisher–Yates shuffling \cite{yeom2020methodology}. However, this has limited effect with messages carrying single-value signals (e.g., all 0s). S2-CAN \cite{pese2021s2} proposed using both insertion and shuffling. Privacy-preserving transformations were proposed by the authors of SIFT \cite{enev2012sensorsift, enev2016automobile} for sensor data to hide privacy-revealing information, but they were never analyzed for CAN (\val: $\Circle$).\looseness-1


\mysub{Protection: Controller Layer}
\setcounter{paranum}{1}

\Dconfsub{ID Anonymization} 
\label{subsubsec:IDObf}
Leaving the ID of a message as is, encryption (\DCa) alone may not be able to completely preserve the confidentiality of a message and defend against confidentiality (\Ca) attacks. Similarly, lack of ID secrecy makes possible attacks that rely on simultaneous-transmission (\Cc, \Ie, \Ij, \Ac-\Ad), since they rely on sending a message with the same ID as a victim message simultaneously. However, since IDs play a pivotal role in priority, schedulability, and filtering, simply encrypting the ID field is problematic for CAN as it disrupts these aspects. To preserve priority and schedulability specifically, researchers proposed using priority bounds, pre-calculated tables, order-preserving one-way functions to anonymize the ID field \cite{lukasiewycz2016security, groza2020highly, brown2020dynamic, cheng2020caneleon, wu2018idh}. However, earlier works \cite{lukasiewycz2016security} have shown several priority-preserving anonymization schemes reverse-engineerable. CanSafe \cite{roy2022cansafe} proposed hardware-level dissociation between IDs and priorities, but without analyzing its impact on schedulability (\infra: \low, \val: \Circle). Other schemes split the ID into a fixed priority field and a dynamic field, but the strict correlation between the new priority field and old IDs makes them easily reverse-engineered. This limits their suitability to authentication and error handling attacks \cite{woo2019can, han2015practical, bhatia2021evading}, but not confidentiality.

\mysub{Protection: Transwire Layer}
\setcounter{paranum}{2}

\Dconfsub{Shielding and Signal Slope Control}\label{subsubsec:shielding} Little work has been done (\val: $\Circle$) to secure CAN from emission-based frame sniffing (\Ce). However, research suggests that shielding and slope control to make the transition edges between dominant and recessive voltages less sharp \cite{silva1999emi} may reduce emissions.\as{[\ins: P, \cov: $--$; \ecu: $--$; \bus: $--$; \infra: $\CIRCLE$; \val: $\Circle$]}

\begin{tcolorbox}[
 colback=blue!5,
  colframe=black!40,
  fonttitle=\bfseries,
  boxsep=0pt,
  left=2pt,
  right=2pt,
  top=2pt,
  bottom=2pt,
  before skip=4pt,
  after skip=4pt,
]
\shortsectionBf{Confidentiality Takeaway:} Weakly defended and restricted in coverage. No highly replicable confidentiality attack has an effective defense, and only one lower-replicability attack is effectively defended. Existing defenses have coverage issues, leaving known attack variants or attacker models unaddressed.

\end{tcolorbox}


\subsection{Integrity Defenses}
\label{subsec:DefenseI}

\mysub{Protection: Higher Layer}
\setcounter{paranum}{0}

\begin{figure}[t]
  \centering
  \includegraphics[width=\linewidth]{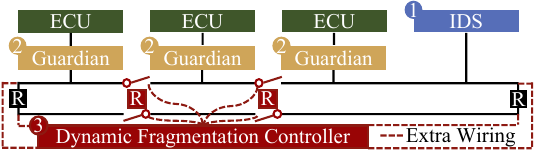}
  \caption{Three defense types and their infrastructure needs.}
  \label{fig:IDS}
\end{figure}

\Dintsub{Content-Inspection}\label{subsubsec:content_inspect}
To detect messages containing fake or malicious data (\Ia, \Id, and \Aa), researchers proposed inspecting message contents. The inspector is deployed as a central IDS (Fig. \ref{fig:IDS}-{\small{\CircledText{1}}}), although some are deployed with individual ECUs (Fig. \ref{fig:IDS}-{\small{\CircledText{2}}}). The inspected content may include message size, ID, DLC, payload range, correlation with other messages and data sources, consistency with previous messages \cite{muter2010structured}, or entropy \cite{muter2011entropy,marchetti2016evaluation}. Such methods have been proposed for other buses as well \cite{eckhardt2019system, eckhardt2020bus}. Inspection uses several machine
learning algorithms \cite{taylor2016anomaly, longari2020cannolo, shahriar2022canshield, kang2016intrusion, kang2016novel, song2020vehicle, hossain2020lstm,shahriarcantropy} sometimes in combination with rule-based checking \cite{tariq2020can}. Some inspectors are made specifically for higher layer protocols (e.g., J1939) \cite{mukherjee2017injection}. Detecting whether the communicated signal value (e.g., of a sensor) is plausible often requires understanding the physical model of the entire system in which CAN is operating (e.g., a car). To do so, researchers proposed several approaches. Some works focused on using heterogeneous sensor consistency \cite{ganesan2017exploiting,guo2019detecting}. Some attempted to understand the context or the state in which the vehicle operates \cite{chen2022context, wasicek2017context, chen2024context, xue2022said}. While these solutions achieved various levels of success, full system modeling
remains a difficult problem with many variables (\cov: $\LEFTcircle$).\looseness-1

\Dintsub{Better Implementation \& Patching}
\label{subsubsec:ProtocolCheck}
Several attacks exploiting implementation problems could be {prevented} (\cov: $\CIRCLE$) by taking measures not impacting the ECU or bus operation.\as{[\ins: P,  \ecu: $\Circle$; \bus: $\Circle$; \infra: $\Circle$]} Checking sequence numbers for validity when handling multi-frame sessions~\cite{muter2010structured} prevents attacks exploiting weak sequence number checking (\Cb, \Ib). Ignoring invalid protocol requests or timing out silent sessions~\cite{chatterjee2021exploiting} prevents attacks making invalid requests (\Aa) or establishing fake sessions (\Ab). Implementing strong diagnostic authentication by taking measures such as avoiding repetitive keys, using stronger authentication algorithms, and rate-limiting unsuccessful authentication attempts can prevent authentication bypass (\Ic) and other attacks that rely on it (\Cb, \Ib, \Aa). Verifying only authenticated devices could send ECU-disabling commands (\Aa) or higher-layer address claims (\Ab) protects availability.

\Dintsub{ECU Firmware Hardening}\label{subsubsec:ecu_hardening} These approaches {prevent} attacks targeting ECUs' firmware (\cov: \CIRCLE). {Firmware obfuscation} \cite{yu2015automobile,milburn2018there} {prevents} attacks revealing firmware or memory (\Cb). Although not proposed or analyzed specifically for CAN, they are used in various systems with typically high computation \cite{cyr2019low,stefanov2018path}. {Firmware attestation} \cite{van2017vulcan,khodari2019decentralized} or {control flow integrity} \cite{nasahl2021protecting,wang2024insectacide} {prevents} attacks manipulating ECUs' firmware integrity or execution flow (\Ib). They often require extra computation (\ecu: \moderate) and special hardware such as hardware security modules (\infra: \moderate). Many are not proposed or evaluated specifically for CAN (\val: \Circle $\rightarrow$ \LEFTcircle).

\begin{tcolorbox}[
 colback=blue!5,
  colframe=black!40,
  fonttitle=\bfseries,
  boxsep=0pt,
  left=2pt,
  right=2pt,
  top=2pt,
  bottom=2pt,
  before skip=4pt,
  after skip=4pt,
]
\shortsectionBf{Higher Layer Takeaway:} Defended in volume, exposed in priority. Most higher-layer attacks are effectively defended, but its highly replicable attacks are not.
\end{tcolorbox}

\mysub{Protection: Controller Layer}
\setcounter{paranum}{3}

\Dintsub{Message Authentication} 
\label{subsubsec:msgAuth}  To prevent {impersonation and replay} attacks (\Id), cryptographic authentication faces challenges discussed in Sec. \ref{subsec:defense_rc}. To overcome these challenges, researchers proposed changing the {authentication immediacy}, {authentication channel}, and {authentication responsibility}: 

\shortsectionEmph{Authentication Channel.} Most solutions embed authentication and freshness data in the payload. If the transmitted data length is short, this would force the message to be longer. However, since the maximum length of a CAN message is already short, this usually requires splitting the data and authentication information across several messages, increasing busload (\bus: \low $\rightarrow$ \moderate) \cite{bella2019toucan, hartkopp2012macan, nurnberger2016vatican}. To tackle the message-length challenge, others suggested using unused message fields \cite{pese2021s2} or the ID \cite{radu2016leia}. Another channel is out-of-band signals or modified CAN fields using special equipment (\infra: \low), such as the CRC \cite{woo2014practical}. CANAuth proposed using CAN+ \cite{van2011canauth, ziermann2009can+}, a modified version of CAN requiring special controllers. CAN-HG \cite{CAN-HG} uses special hardware to alter the baud rate within the frame \cite{tindelldefending}, allowing the embedding of authentication data. Timing channels are discussed separately in \DIh.

\shortsectionEmph{Authentication Immediacy.} \textit{Immediate authentication} communicates the authentication information by the end of the message \cite{kurachi2014cacan, wang2017hardware, serag2023zbcan, bella2019toucan, hartkopp2012macan}. However, the short message length limits the amount of both payload and authentication data communicated in a message. To tackle the length challenge, \textit{delayed authentication} waits for a subsequent message {(\bus: \low)} to authenticate retrospectively \cite{nurnberger2016vatican}. An example is the TESLA protocol, which uses one-way chains \cite{perrig2003tesla}. Groza and Murvay developed and later refined a centralized version of it for CAN \cite{groza2011higher, groza2013efficient}.

\shortsectionEmph{Authentication Responsibility.} In \textit{per-ECU authentication}, each receiver authenticates incoming messages \textit{(\ecu: \low, \ins: P, D}). However, this may impact the ECU performance as ECUs are typically resource-constrained. In \textit{centralized authentication}, a monitor node authenticates messages instead of every ECU \textit{(\ins: D)}, eliminating the overhead of authenticating received messages \textit{(\ecu: \moderate)}. The monitor is installed in a similar way to an IDS in Fig. \ref{fig:IDS}-{\small{\CircledText{1}}}, to prevent it from being a bottleneck. It has different keys for each ECU, alleviating the internal-attacker/group-key management issue (discussed in Sec. \ref{subsec:defense_rc}) {(\cov: $\CIRCLE$)}. Nodes embed the authentication data in the {authentication channel} of outgoing messages \cite{jo2019mauth, kurachi2014cacan, wang2017hardware,wagner2025caiba}. If {immediate authentication} is in use, the monitor could be used with {frame destruction} (\DIe) \cite{kurachi2014cacan, wang2017hardware} to prevent malicious frames from being received since this approach has a very low false positive rate (\ins: P*).\as{ [\ins: D; \cov: $\CIRCLE$; \ecu: $\LEFTcircle$; \bus: $\LEFTcircle$; \infra: $\Circle$; \val: $\CIRCLE$]} 

\shortsectionEmph{Diagnostic Sessions.} Worth more consideration due to their range of control, the {source}, {content}, and {freshness} should be authenticated using {per-ECU} keys without group keys. To prevent obtaining authentication keys through brute-forcing or reverse engineering, OBD-II dongles limiting the number of attempts within a time window, stronger keys, or multi-factor authentication could be used. Assuming diagnostic sessions are infrequent and do not happen during normal operation, any authentication channel could be used without impacting the performance of the bus or ECUs.\as{ [I: P; C: $\CIRCLE$; E: $\Circle$; B: $\Circle$; F: $\Circle$; V: $\CIRCLE$]}


\Dintsub{Frame Destruction} 
\label{subsubsec:FrameDest} Injecting an error frame or overwriting a 1 with a 0 during frame transmission interrupts it and causes receiving CAN controllers to discard it without passing it to the higher layer. This technique must be combined with an IDS or a centralized authentication approach, enabling an IDS capable of injecting dominant bits to destroy the frames they deem malicious (\ins: P*, \cov: $\CIRCLE$)\cite{giannopoulos2015can, serag2023zbcan, tindelldefending, kurachi2014cacan, wang2017hardware}.\looseness-1


\Dintsub{Victim Resistance}
\label{subsubsec:victimRes} Several works proposed that every ECU monitor whether they see their IDs transmitted on the bus by someone else and react upon detecting an impersonating message (\Id) \cite{TJA115x, matsumoto2012method, dagan2016parrot, pese2025michican}. The most effective approach is the secure transceiver. It is a dedicated hardware device that monitors the IDs autonomously without ECU software processing. It raises an error upon seeing an ID that belongs to its ECU, effectively employing frame destruction (\DIe) but in a decentralized manner \cite{TJA115x}.\as{[\ins: P; \cov: $\CIRCLE$; \ecu: $\Circle$; \bus: $\Circle$; \infra: $\LEFTcircle$; \val: $\CIRCLE$]} 


\Dintsub{Physical IDS (PIDS)}
\label{subsubsec:IDS}
\label{subsubsec:VIDS}
Monitoring signal characteristics may identify message source or transmitter locations. When compared with the expected source or location, it could identify anomalies or reveal the source ECU of an attack. The IDS is usually placed on the bus in parallel (Fig. \ref{fig:IDS}-{\small{\CircledText{1}}}) so it does not constitute a bottleneck. Accumulation of research led to solutions with low false-positive rates that may be combined with {frame-destruction} (\DIe) to {prevent} {impersonation \& replay} (\Id). Researchers have proposed various IDS types that monitor physical signal characteristics: 

\shortsectionEmph{Voltage and Current-Based IDS (VIDS)}. To detect {impersonation and replay attacks}, works profiled voltage characteristics of different nodes using a voltage sampling device \cite{choi2018voltageids, murvay2014source, cho2017viden, choi2018identifying, kneib2020easi, xun2021vehicleeids, kneib2018scission, foruhandeh2019simple, murvay2020tidal, levy2021can}. It compares a message's voltage profile to its legitimate transmitter to detect impersonation. Others extended VIDS to detect certain {by-the-rules} error handling attacks (\Ac-\Ad) \cite{choi2018voltageids}. Due to environmental variations, voltage profiles change over time and thus need to be updated with newly collected measurements. These online updates present fertile ground for training set poisoning attacks (i.e., attackers manipulating voltage measurements to slowly corrupt the profiles learned by the IDS) if they are not secured \cite{bhatia2021evading}. Others noticed that the electrical current drawn when sending a 1 bit by a single transmitter is different from that drawn when the bit was 0 but has been flipped by another attacker, and proposed using that to detect {frame tampering} attacks that flip 0 to 1 (\Ii) \cite{rogers2022detecting}.\as{ [\ins: D; \cov: $\LEFTcircle$; \ecu: $\Circle$; \bus: $\Circle$; \infra: $\Circle$]} 

\shortsectionEmph{Frequency and Time-Based IDS (TIDS).}
\label{subsubsec:TIDS}
Researchers noticed that the transmission frequency of various CAN messages with different IDs \cite{taylor2015frequency, song2016, hoppe2011security, Young2019Frequency}, or the timing \cite{cho2016fingerprinting, olufowobi2019saiducant, lee2017otids, tang2025wip} or sequence patterns \cite{marchetti2017anomaly, tomlinson2018detection, katragadda2020detecting} of periodic or request-response messages are largely fixed under normal operation and proposed monitoring them to detect anomalies. Others monitored timing features within the frame, such as bit-width, rise and fall times, and asymmetry \cite{zhou2019btmonitor,ohira2020divider,ohira2021pli,schell2022asymmetric, tang2024eracan} as they are largely fixed for the same transmitter.\looseness-1 

\label{subsubsec:localize}
\shortsectionEmph{Attacker Localization.} Detecting transmitter (attacker) location is possible using wires connecting both ends of the bus to a TDC \cite{roeschlin2023edgetdc}, or a voltage sampler \cite{levy2021can, murvay2020tidal, groza2021can}. Monitoring the difference in message arrival time between the two ends allows localizing the transmitter with an accuracy of up to 10 cm. These approaches work against all attacker types, including those with physical access.

\Dintsub{Unifying Sampling Points}\label{subsubsec:unify_sp} Unifying the voltage sample points of all controllers on the bus could prevent attacks that exploit sample point discrepancies to make different controllers read the same message differently. These include {poly-semantic frames} (\Ih) and {frame tampering} (\Ii) attacks.\as{ [\ins: P; \cov: $\CIRCLE$; \ecu: $\Circle$; \bus: $\Circle$; \infra: $\LEFTcircle$; \val: $\LEFTcircle$]}

\mysub{Protection: Transwire Layer}
\setcounter{paranum}{8}

\Dintsub{Time and IFS Anonymization}
\label{subsubsec:zbcan}
\label{subsubsec:IFSAuth}Researchers proposed to encode secrets in timing channels (i.e., sending the message after a secret delay or at a secret time) to achieve authentication and protection against attacks targeting messages with predictable transmission times \cite{groza2018incanta, groza2020canto, ying2019tacan, vanderhallen2021robust}. However, because CAN's priority system can delay messages unpredictably, precisely controlling their absolute transmission time to encode secrets was not robust. These solutions also heavily impacted schedulability and incurred high false positive rates. \MR{8} \mbox{ZBCAN} \cite{serag2023zbcan} proposed eliminating the schedulability and false-positive issues using IFS (time since the last message terminated on the bus) instead of absolute message time as a robust channel to embed secrets in combination with frame destruction (\DIe) to prevent {impersonation}. It can also work against error-handling attacks targeting messages with predictable transmission times (i.e., simultaneous-transmission-based: \Cc, \Ie, \Ij, and \Ac-\Ad) when launched by {by-the-rule}s attackers, but it fails against {bit-bangers}.\looseness-1

\Dintsub{Bit-Banger IDS}
\label{subsubsec:layerIIIDS} 
Since {bit-banger} attacks typically manipulate lower-layer events, they cannot be detected by monitoring message-level features (e.g., transmission frequency). {Bit-banger} IDSs monitor the link-layer at the bit level and physical layer features within frames to detect signatures that correspond to {bit-banger} attacks. This detects attacks including \Id-\Ij~and \Ac-\Af. \MR{8}ERACAN \cite{tang2024eracan} proposed that a central IDS attach to the bus (\infra: \high), watch it, and use {frame destruction} to prevent attacks such as \Id-\Ie~and \Ig-\Ii. CANTXSec proposed that the IDS be connected to each ECU's CAN TX pins to identify the attacker among them (\infra: \low) \cite{donadel2025cantxsec}.


\Dintsub{ECU Guardian}
\label{subsubsec:guardian}
\label{subsubsec:j1939firewall} Researchers proposed physically installing a guardian between each ECU and the bus (Fig. \ref{fig:IDS}-{\small{\CircledText{2}}}, \infra: \low). This could both protect the ECU from attacks from the bus and the bus from attacks from the ECU. This includes most attacks, including attacks on higher-layer protocols (e.g., J1939) \cite{Daily2021SecuringCT}.\as{[\ins: P; \ecu: $\Circle$; \bus: $\Circle$; \infra: \low]} In CAN-HG, the guardian could detect malicious activities itself or in collaboration with a central IDS. It can physically disconnect the ECU on detecting malicious activities by the guardian or receiving an out-of-band IDS signal \cite{Hijack, tindelldefending}.\looseness-1

\begin{tcolorbox}[
 colback=blue!5,
  colframe=black!40,
  fonttitle=\bfseries,
  boxsep=0pt,
  left=2pt,
  right=2pt,
  top=2pt,
  bottom=2pt,
  before skip=4pt,
  after skip=4pt,
]
\shortsectionBf{Integrity Takeaway:} Strong attack--defense convergence. Integrity is both the most targeted CIA element and the most defended: 88 of 107 defense papers focus on it. This effectively defends 7 of its 10 attacks, including 4 of 6 highly replicable attacks, making integrity the only CIA element with effective defenses for highly replicable attacks.

\end{tcolorbox}




\subsection{Availability Defenses}
\label{subsec:DefenseA}



\mysub{Protection: Controller Layer}
\setcounter{paranum}{0}



\Davsub{Error Handling IDS}\label{subsubsec:err_ids} Researchers proposed installing a central IDS that monitors errors on the bus and models the error counters of various ECUs to detect when they transition into the bus-off state \cite{longari2019copycan}. They did not discuss other states such as the error-passive state, but the same method can likely detect it in the same way, thereby detecting any attack pushing the victim into specific error states (\Cc, \Ie, \Ij, and \Ad).

\Davsub{Layer-II Node Suspension}
\label{subsubsec:nodeSuspend}
Since nodes stop communicating in the bus-off state, researchers proposed suspending {by-the-rules} attackers to the bus-off state to stop ongoing attacks. For attacks targeting availability, such as those attacking the messages of other ECUs or denying them bus access (\Ac-\Ae), this gives room for other nodes to communicate. The faster the attacker is suspended, the more effective the defense, especially for \Ae~(\MR{8}Several techniques explained in \Ad). However, it may be dangerous if the attacking ECU is safety-critical with only some malicious messages (e.g., if the brake ECU is attacking the infotainment ECU, it would not make sense to suspend the brake ECU). Further, it does not work against {bit-banger}s or {down-to-earther}s, as they do not abide by layer-II rules, including transitioning to the bus-off state. \as{[\ins: P; \cov: $\LEFTcircle$; \ecu: $\Circle$; \bus: $\Circle$; \infra: $\Circle$; \val: $\CIRCLE$]}

\Davsub{ID-Priority Dissociation}\label{subsubsec:id_prio_dis} Since the priority of accessing the bus in CAN is tied to the CAN ID, CANSafe proposed a complete change of CAN's physical layer (\infra: \low) to dissociate ID and priority \cite{roy2022cansafe}. This could prevent attackers from flooding the bus with high-priority IDs to cause other message IDs to starve (\Ae). However, although priority affects schedulability, the authors did not analyze the impact of such a change on it.

\Davsub{One-Shot mode} \label{subsubsec:oneShot}
Most CAN controllers offer disabling automatic retransmission (called one-shot transmission mode in several controllers) as an optional mode \cite{microchip2007controller,philips2000controller}. This cheap defense does not require high operational or infrastructure costs to {slow down} or even {prevent} many {by-the-rules} (\cov: $\LEFTcircle$) attacks that rely on pushing the victim ECU to specific error states (\Cc, \Ie, \Ij, and \Ad). This is because many such attacks rely on repetitively attacking retransmitted messages to increase the victim ECU's error counter and push it into the desired error state. However, it has little effectiveness against non-retransmission-based attacks.

\begin{tcolorbox}[
  colback=pink!15,
  colframe=pink!60,
  fonttitle=\bfseries,
  boxsep=0pt,
  left=2pt,
  right=2pt,
  top=2pt,
  bottom=2pt,
  before skip=4pt,
  after skip=4pt,
]
\textbf{Observation:} All 4 attacks targeting controller layer availability are highly replicable and have no effective defenses. 
\end{tcolorbox}

\begin{tcolorbox}[
 colback=blue!5,
  colframe=black!40,
  fonttitle=\bfseries,
  boxsep=0pt,
  left=2pt,
  right=2pt,
  top=2pt,
  bottom=2pt,
  before skip=4pt,
  after skip=4pt,
]
\shortsectionBf{Controller Layer Takeaway:} Heavily studied yet exposed. Although 70 of 107 defense papers address controller-layer attacks, most attacks still lack effective defenses. Existing solutions either lack prevention, especially against {bit-bangers}, or require high infrastructure changes.

\end{tcolorbox}


\mysub{Protection: Transwire Layer}
\setcounter{paranum}{4}


\Davsub{Dynamic Fragmentation} \label{subsubsec:fragmentation} Researchers suggested modifying the bus wiring to support attacker isolation, especially those targeting availability (e.g., \Ae) \cite{groza2021canary, wolf2004security}. CANARY installs a central IDS and relays at specific locations on the bus (Fig. \ref{fig:IDS}-{\small{\CircledText{3}}}, \infra: \low). When the IDS detects an attack, CANARY uses the relays to disconnect each section of the bus one by one, search for the section where attacks are still observed to localize the attacker, and disconnect the section containing the attacker. However, this setup could result in isolating entire sections of the bus and not only the attacker. While its infrastructure cost and response time are relatively high, it offers the only line of defense against attacks with no other countermeasures and could be generalized to {prevent} any attack with an effective detection strategy (\ins: P*).\as{[\ins: P*; \cov: $\CIRCLE$; \ecu: $\Circle$; \bus: $\Circle$; \infra: $\CIRCLE$]}

\begin{tcolorbox}[
 colback=blue!5,
  colframe=black!40,
  fonttitle=\bfseries,
  boxsep=0pt,
  left=2pt,
  right=2pt,
  top=2pt,
  bottom=2pt,
  before skip=4pt,
  after skip=4pt,
]
\shortsectionBf{Transwire Layer Takeaway:} Covered where it matters. The transwire layer is the mirror image of the higher layer. It receives the least defense attention (17 of 107 defense papers) and has only one effective defense. Yet that defense covers the layer’s only highly replicable attack (\Ij).

\end{tcolorbox}

\begin{tcolorbox}[
 colback=blue!5,
  colframe=black!40,
  fonttitle=\bfseries,
  boxsep=0pt,
  left=2pt,
  right=2pt,
  top=2pt,
  bottom=2pt,
  before skip=4pt,
  after skip=4pt,
]
\shortsectionBf{Availability Takeaway:} Underdefended yet over-targeted. Availability is the second most targeted CIA element, but the least studied in defenses: only 5 of 107 papers focus on it. All its highly replicable {bit-banger} attacks lack effective defenses. Existing defenses either lack prevention or require high infrastructure changes.
\end{tcolorbox}



\section{Root Cause Analysis}
\label{A:Analysis}
\label{A:ResearchGaps}

To determine whether CAN security problems are fundamentally CAN-specific, we analyze the root causes underlying its attacks and defense gaps. For each root cause, we determine whether it is \textit{inherent} to CAN, \textit{unique} to CAN, or both, and use these classifications to infer which attacks and gaps are themselves inherent or unique. This distinguishes problems rooted in CAN's design from those arising from broader shared-media properties, higher-layer protocols and implementations, or the surrounding system. Sec.~\ref{sec:investigat} builds on these root causes to investigate whether the same problems are likely to persist in emerging IVBs.

Our analysis identifies 9 attack root causes ($\mathcal{R}$) and 3 defense-gap root causes ($\mathcal{DR}$). This section describes and analyzes each root cause along two non-mutually-exclusive qualities: \textit{1) Inherency (I):} inherent to the standard or attacker model, rather than accidental or scenario-specific. \textit{2) Uniqueness (U):} specific to the CAN family of standards. Because a root cause may underlie multiple attacks or gaps, we organize the analysis by root cause and identify the attacks or gaps each one underlies. Tbl.~\ref{tbl:root_cause} summarizes the root-cause analysis. It maps each attack and defense gap to its underlying root causes and reports whether each root cause is inherent, unique, or both. A \checkmark~indicates that a root cause is indispensable to an attack or gap, while an $\uparrow$ indicates that addressing an attack root cause provides an effective defense. By-the-rules gaps are highlighted in red, bit-banger gaps in blue, and excluded attacks and root causes in gray.

\definecolor{Matrix}{rgb}{0.65,0.301,0.301}
\definecolor{BeautyBush}{rgb}{0.90,0.70,0.60}
\definecolor{Manhattan}{rgb}{0.7600,0.7955,0.8992}
\definecolor{BurntOrange}{rgb}{0.3373, 0.4353, 0.7216}
\definecolor{Bobo}{RGB}{176,145,122}
\definecolor{SlateGray}{rgb}{0.78, 0.78, 0.80}   

\begin{table}[t]
\caption{\MR{6}Root Causes for CAN Attacks and Gaps} 
\label{tbl:root_cause}
  \centering
  { \setlength{\tabcolsep}{2pt}
  \resizebox{\columnwidth}{!}{
  \begin{tabular}{||l||c|c|c|c|c|c|c|c|c|c|c|c||c|c|c|c|c||}
\hhline{~|t:============:t:=====:t|}
\multicolumn{1}{c||}{}
& {\rotatebox{90}{\parbox[b]{8em}{\raggedright$\mathcal R1\,(I)$}}}
& {\rotatebox{90}{\parbox[b]{8em}{\raggedright$\mathcal R2\,(I)$}}}
& {\rotatebox{90}{\parbox[b]{8em}{\raggedright$\mathcal R3\,(I)$}}}
& {\rotatebox{90}{\parbox[b]{8em}{\raggedright$\mathcal R4\,(IU)$}}}
& {\rotatebox{90}{\parbox[b]{8em}{\raggedright$\mathcal R5\,(I)$}}}
& {\rotatebox{90}{\parbox[b]{8em}{\raggedright$\mathcal R6\,(IU)$}}}
& {\rotatebox{90}{\parbox[b]{8em}{\raggedright$\mathcal R7$}}}
& {\rotatebox{90}{\parbox[b]{8em}{\cellcolor{SlateGray}\raggedright$\mathcal R8$}}}
& {\rotatebox{90}{\parbox[b]{8em}{\cellcolor{SlateGray}\raggedright$\mathcal R9$}}}
& {\rotatebox{90}{\parbox[b]{8em}{\raggedright$\mathcal{DR}1\,(I)$}}}
& {\rotatebox{90}{\parbox[b]{8em}{\raggedright$\mathcal{DR}2\,(I)$}}}
& {\rotatebox{90}{\parbox[b]{8em}{\raggedright$\mathcal{DR}3$}}}
& {\rotatebox{90}{\parbox[b]{8em}{\raggedright CAN FD}}}
& {\rotatebox{90}{\parbox[b]{8em}{\raggedright CAN XL (Err. On)}}}
& {\rotatebox{90}{\parbox[b]{8em}{\raggedright CAN XL (Err. Off)}}}
& {\rotatebox{90}{\parbox[b]{8em}{\raggedright 10BASE-T1S}}}
& {\rotatebox{90}{\parbox[b]{8em}{\raggedright Ideal Bus}}}
\\
\hhline{|t:=::============::=====:|}
\cellcolor{Matrix}\textcolor{white}{\Ca. Communication Surveil} &\checkmark &\checkmark & & & & & &\cellcolor{SlateGray} &\cellcolor{SlateGray} &\checkmark & & & \cellcolor{Matrix}$\CIRCLE$ & \cellcolor{Matrix}$\CIRCLE$ & \cellcolor{Matrix}$\CIRCLE$ & \cellcolor{Matrix}$\CIRCLE$ & $\Circle$\\
\hhline{||-||------------||-----||}
\Cb. Memory Leakage &\checkmark & & & & & &\checkmark &\cellcolor{SlateGray} &\cellcolor{SlateGray} & & & & $\CIRCLE$ & $\CIRCLE$ & $\CIRCLE$ & $\CIRCLE$ & $\CIRCLE$\\
\hhline{||-||------------||-----||}
\cellcolor{BurntOrange}\textcolor{white}{\Cc. Network Reconnaissance} &\checkmark &\checkmark &\checkmark &\checkmark &\checkmark & & &\cellcolor{SlateGray} &\cellcolor{SlateGray} & &\checkmark & & $\CIRCLE$ & $\CIRCLE$ & - & \cellcolor{Matrix}$\CIRCLE$ & $\Circle$\\
\hhline{||-||------------||-----||}
\cellcolor{Matrix}\textcolor{white}{\Ia. Data Fabrication} &\checkmark & & & & & & &\cellcolor{SlateGray} &\cellcolor{SlateGray} & & &\checkmark & \cellcolor{Matrix}$\CIRCLE$ & \cellcolor{Matrix}$\CIRCLE$ & \cellcolor{Matrix}$\CIRCLE$ & \cellcolor{Matrix}$\CIRCLE$ & \cellcolor{Matrix}$\CIRCLE$\\
\hhline{||-||------------||-----||}
\Ib. Mem \& FW Manipulation &\checkmark & & & & & &\checkmark &\cellcolor{SlateGray} &\cellcolor{SlateGray} & & & & $\CIRCLE$ & $\CIRCLE$ & $\CIRCLE$ & $\CIRCLE$ & $\CIRCLE$\\
\hhline{||-||------------||-----||}
\Ic. Bypass Diagnostic Auth. &\checkmark & & & & & &\checkmark &\cellcolor{SlateGray} &\cellcolor{SlateGray} & $\uparrow$ & & & $\CIRCLE$ & $\CIRCLE$ & $\CIRCLE$ & $\CIRCLE$ & $\CIRCLE$\\
\hhline{||-||------------||-----||}
\Id. Impersonation \& Replay &\checkmark &\checkmark & & & & & &\cellcolor{SlateGray} &\cellcolor{SlateGray} &$\uparrow$ & & & $\CIRCLE$ & $\CIRCLE$ & $\CIRCLE$ & \cellcolor{Matrix}$\CIRCLE$ & $\Circle$\\
\hhline{||-||------------||-----||}
\Ie. Frame Hijack &\checkmark &\checkmark &\checkmark &\checkmark &\checkmark & & &\cellcolor{SlateGray} &\cellcolor{SlateGray} & & &$\uparrow$ & $\CIRCLE$ & $\CIRCLE$ & - & - & $\Circle$\\
\hhline{||-||------------||-----||}
\Ij. Fingerprint Distortion &\checkmark &\checkmark &\checkmark &\checkmark &\checkmark & & &\cellcolor{SlateGray} &\cellcolor{SlateGray} & &$\uparrow$ & & $\CIRCLE$ & $\CIRCLE$ & $\CIRCLE$ & - & $\Circle$\\
\hhline{||-||------------||-----||}
\Aa. ECU Disabling &\checkmark & & & & & &\checkmark &\cellcolor{SlateGray} &\cellcolor{SlateGray} & & & & $\CIRCLE$ & $\CIRCLE$ & $\CIRCLE$ & $\CIRCLE$ & $\CIRCLE$\\
\hhline{||-||------------||-----||}
\Ab. Sybil Exhaustion &\checkmark & & & & & &\checkmark &\cellcolor{SlateGray} &\cellcolor{SlateGray} & & & & $\CIRCLE$ & $\CIRCLE$ & $\CIRCLE$ & $\CIRCLE$ & $\CIRCLE$\\
\hhline{||-||------------||-----||}
\cellcolor{BurntOrange}\textcolor{white}{\Ac. Error Injection} &\checkmark &\checkmark &\checkmark &\checkmark &\checkmark & & &\cellcolor{SlateGray} &\cellcolor{SlateGray} & &\checkmark & & $\CIRCLE$ & $\CIRCLE$ & $\CIRCLE$ & - & $\Circle$\\
\hhline{||-||------------||-----||}
\cellcolor{BurntOrange}\textcolor{white}{\Ad. Error State Manipulation} &\checkmark &\checkmark &\checkmark &\checkmark &\checkmark & & &\cellcolor{SlateGray} &\cellcolor{SlateGray} & &\checkmark & & $\CIRCLE$ & $\CIRCLE$ & $\Circle$ & $\Circle$ & $\Circle$\\
\hhline{||-||------------||-----||}
\cellcolor{BurntOrange}\textcolor{white}{\Ae. Bus Access Denial} &\checkmark &\checkmark &\checkmark & & & & &\cellcolor{SlateGray} &\cellcolor{SlateGray} & &\checkmark & & $\CIRCLE$ & \cellcolor{Matrix}$\CIRCLE$ & \cellcolor{Matrix}$\CIRCLE$ & \cellcolor{Matrix}$\CIRCLE$ & $\Circle$\\
\hhline{|:=::============::=====:b|}
\cellcolor{SlateGray}\Cd. HW \& Topology Profile
  &\cellcolor{SlateGray}\checkmark &\cellcolor{SlateGray}\checkmark
  &\cellcolor{SlateGray} &\cellcolor{SlateGray} &\cellcolor{SlateGray}\checkmark
  &\cellcolor{SlateGray} &\cellcolor{SlateGray}
  &\cellcolor{SlateGray} &\cellcolor{SlateGray}\checkmark
  &\cellcolor{SlateGray} &\cellcolor{SlateGray} &\cellcolor{SlateGray}
  &\multicolumn{1}{c}{} &\multicolumn{1}{c}{} &\multicolumn{1}{c}{} &\multicolumn{1}{c}{} &\multicolumn{1}{c}{} \\
\hhline{||-||------------||}
\cellcolor{SlateGray}\Ce. EM Emission
  &\cellcolor{SlateGray}\checkmark &\cellcolor{SlateGray}
  &\cellcolor{SlateGray} &\cellcolor{SlateGray} &\cellcolor{SlateGray}\checkmark
  &\cellcolor{SlateGray} &\cellcolor{SlateGray}
  &\cellcolor{SlateGray} &\cellcolor{SlateGray}\checkmark
  &\cellcolor{SlateGray} &\cellcolor{SlateGray} &\cellcolor{SlateGray}
  &\multicolumn{1}{c}{} &\multicolumn{1}{c}{} &\multicolumn{1}{c}{} &\multicolumn{1}{c}{} &\multicolumn{1}{c}{} \\
\hhline{||-||------------||}
\cellcolor{BurntOrange}\textcolor{white}{\IF. Double Receive}
  &\cellcolor{SlateGray}\checkmark &\cellcolor{SlateGray}\checkmark
  &\cellcolor{SlateGray} &\cellcolor{SlateGray}\checkmark &\cellcolor{SlateGray}\checkmark
  &\cellcolor{SlateGray} &\cellcolor{SlateGray}
  &\cellcolor{SlateGray}\checkmark &\cellcolor{SlateGray}
  &\cellcolor{SlateGray} &\cellcolor{SlateGray}\checkmark &\cellcolor{SlateGray}
  &\multicolumn{1}{c}{} &\multicolumn{1}{c}{} &\multicolumn{1}{c}{} &\multicolumn{1}{c}{} &\multicolumn{1}{c}{} \\
\hhline{||-||------------||}
\cellcolor{SlateGray}\Ig. Unorthodox Frames
  &\cellcolor{SlateGray}\checkmark &\cellcolor{SlateGray}
  &\cellcolor{SlateGray} &\cellcolor{SlateGray}\checkmark &\cellcolor{SlateGray}
  &\cellcolor{SlateGray} &\cellcolor{SlateGray}
  &\cellcolor{SlateGray}\checkmark &\cellcolor{SlateGray}
  &\cellcolor{SlateGray} &\cellcolor{SlateGray} &\cellcolor{SlateGray}
  &\multicolumn{1}{c}{} &\multicolumn{1}{c}{} &\multicolumn{1}{c}{} &\multicolumn{1}{c}{} &\multicolumn{1}{c}{} \\
\hhline{||-||------------||}
\cellcolor{SlateGray}\Ih. Poly-Semantic Frames
  &\cellcolor{SlateGray}\checkmark &\cellcolor{SlateGray}\checkmark
  &\cellcolor{SlateGray} &\cellcolor{SlateGray} &\cellcolor{SlateGray}
  &\cellcolor{SlateGray}\checkmark &\cellcolor{SlateGray}
  &\cellcolor{SlateGray}\checkmark &\cellcolor{SlateGray}
  &\cellcolor{SlateGray} &\cellcolor{SlateGray} &\cellcolor{SlateGray}
  &\multicolumn{1}{c}{} &\multicolumn{1}{c}{} &\multicolumn{1}{c}{} &\multicolumn{1}{c}{} &\multicolumn{1}{c}{} \\
\hhline{||-||------------||}
\cellcolor{SlateGray}\Ii. Frame Tamper
  &\cellcolor{SlateGray}\checkmark &\cellcolor{SlateGray}\checkmark
  &\cellcolor{SlateGray} &\cellcolor{SlateGray} &\cellcolor{SlateGray}\checkmark
  &\cellcolor{SlateGray} &\cellcolor{SlateGray}
  &\cellcolor{SlateGray}\checkmark &\cellcolor{SlateGray}
  &\cellcolor{SlateGray} &\cellcolor{SlateGray} &\cellcolor{SlateGray}$\uparrow$
  &\multicolumn{1}{c}{} &\multicolumn{1}{c}{} &\multicolumn{1}{c}{} &\multicolumn{1}{c}{} &\multicolumn{1}{c}{} \\
\hhline{||-||------------||}
\cellcolor{BurntOrange}\textcolor{white}{\Af. Synch. Disruption}
  &\cellcolor{SlateGray}\checkmark &\cellcolor{SlateGray}\checkmark
  &\cellcolor{SlateGray} &\cellcolor{SlateGray} &\cellcolor{SlateGray}\checkmark
  &\cellcolor{SlateGray} &\cellcolor{SlateGray}
  &\cellcolor{SlateGray}\checkmark &\cellcolor{SlateGray}
  &\cellcolor{SlateGray} &\cellcolor{SlateGray}\checkmark &\cellcolor{SlateGray}
  &\multicolumn{1}{c}{} &\multicolumn{1}{c}{} &\multicolumn{1}{c}{} &\multicolumn{1}{c}{} &\multicolumn{1}{c}{} \\
\hhline{||-||------------||}
\cellcolor{SlateGray}\Ag. Signal Attenuation
  &\cellcolor{SlateGray}\checkmark &\cellcolor{SlateGray}\checkmark
  &\cellcolor{SlateGray} &\cellcolor{SlateGray} &\cellcolor{SlateGray}\checkmark
  &\cellcolor{SlateGray} &\cellcolor{SlateGray}
  &\cellcolor{SlateGray}\checkmark &\cellcolor{SlateGray}
  &\cellcolor{SlateGray} &\cellcolor{SlateGray}$\uparrow$ &\cellcolor{SlateGray}
  &\multicolumn{1}{c}{} &\multicolumn{1}{c}{} &\multicolumn{1}{c}{} &\multicolumn{1}{c}{} &\multicolumn{1}{c}{} \\
\hhline{||-||------------||}
CAN FD &\CIRCLE &\CIRCLE &\CIRCLE &\CIRCLE &\CIRCLE &\CIRCLE &\CIRCLE &\cellcolor{SlateGray}- &\cellcolor{SlateGray}- & \LEFTcircle &\CIRCLE &\CIRCLE & \multicolumn{1}{c}{} & \multicolumn{1}{c}{} & \multicolumn{1}{c}{} & \multicolumn{1}{c}{} & \multicolumn{1}{c}{} \\
\hhline{||-||------------||~~~~~}
CAN XL (Err. On) &\CIRCLE &\CIRCLE &\CIRCLE &\CIRCLE &\CIRCLE &\CIRCLE &\CIRCLE &\cellcolor{SlateGray}- &\cellcolor{SlateGray}- & \LEFTcircle &\CIRCLE &\CIRCLE & \multicolumn{1}{c}{} & \multicolumn{1}{c}{} & \multicolumn{1}{c}{} & \multicolumn{1}{c}{} & \multicolumn{1}{c}{} \\
\hhline{||-||------------||~~~~~}
CAN XL (Err. Off) &\CIRCLE &\CIRCLE &\CIRCLE &\LEFTcircle &\CIRCLE &\CIRCLE &\CIRCLE &\cellcolor{SlateGray}- &\cellcolor{SlateGray}- & \LEFTcircle &\CIRCLE &\CIRCLE & \multicolumn{1}{c}{} & \multicolumn{1}{c}{} & \multicolumn{1}{c}{} & \multicolumn{1}{c}{} & \multicolumn{1}{c}{} \\
\hhline{||-||------------||~~~~~}
10BASE-T1S &\CIRCLE &\CIRCLE &\CIRCLE &\Circle &\CIRCLE &\Circle &\CIRCLE &\cellcolor{SlateGray}- &\cellcolor{SlateGray}- & \LEFTcircle &\CIRCLE &\CIRCLE & \multicolumn{1}{c}{} & \multicolumn{1}{c}{} & \multicolumn{1}{c}{} & \multicolumn{1}{c}{} & \multicolumn{1}{c}{} \\
\hhline{||-||------------||~~~~~}
Ideal Bus &\CIRCLE &\Circle &\Circle &\Circle &\LEFTcircle &\Circle &\CIRCLE &\cellcolor{SlateGray}- &\cellcolor{SlateGray}- & \Circle &\Circle &\CIRCLE & \multicolumn{1}{c}{} & \multicolumn{1}{c}{} & \multicolumn{1}{c}{} & \multicolumn{1}{c}{} & \multicolumn{1}{c}{} \\
\hhline{|b:=:b:============:b|~~~~~}
  \end{tabular}}}
\end{table}

\subsection{Attack Root Causes ($\mathcal{R}$s)}

\shortsectionEmph{$\mathcal{R}$1: Attacker Access to Network (I).} The mere access of the attacker to the network, whether physically or remotely, is \textit{inherent} to all of our attacker models (Sec. \ref{subsec:BusAccess}) and all attacks require it. While most attacks require additional root causes with it, it is enough for \textit{fake data fabrication} (\Ia). It is {not unique}, as any comparison with other communication systems will assume it under any attacker model. 

\shortsectionEmph{$\mathcal{R}$2: Shared Media Properties (I).} Two {inherent} properties that are {not unique} to CAN, but common to all single-channel shared communication media and exploitable by all our attacker models, underlie many attacks. \textit{1) Broadcast promiscuity:} this refers to the ability of a connected malicious node to anonymously monitor bus-wide activity and to have bus-wide reachability, allowing it to act on any observed frame, node, or the medium itself. This directly enables {communication surveillance}, {network reconnaissance}, and {hardware and topology profiling} attacks. It also underlies attacks that interfere with in-flight frames (\Ie, \Ij, and \Ag), attacks that induce errors or manipulate node state (\Ac~and \Ad), and attacks that exploit discrepancies among nodes sharing the same bus (\Ih, \Af, \IF, \Id). \textit{2) Mutual exclusion:} because transmission on a single shared channel prevents other nodes from transmitting at the same time, it gives an adversary leverage to deny others bus access or seize transmission opportunities, thereby underlying {bus-access denial} and {frame hijacking}.

\shortsectionEmph{$\mathcal{R}$3: Exploitable Priority System (I).} As described in Sec. \ref{sec:background}, CAN uses a bus-wide priority mechanism, physically prioritizing dominant voltage over recessive. On the controller layer, this is utilized for collision resolution using arbitration, where lower CAN IDs (more leading dominant bits) have higher priority. This enables {bus-access denial}, whether by flooding the bus with low-CAN-ID frames or by continuously asserting dominant voltage. It also enables attacks that transmit past arbitration ({network reconnaissance}, {frame hijacking}, {physical fingerprint distortion}, and {error injection and state manipulation}), whether by matching the ID with victim messages to exploit the simultaneous-transmission technique for {by-the-rules} attackers (Sec. \ref{subsubsec:simultaneous}), or by directly flipping recessive voltage to dominant for {bit-bangers}. While this root-cause is {inherent} and exploitable by all attacker models, it is {not unique} as all shared media need priority and media access systems--many of which are exploitable, as shown in Sec. \ref{sec:investigat}.

\shortsectionEmph{$\mathcal{R}$4: Error Handling Properties (IU).} Four {inherent and unique} properties of CAN's error handling mechanism are exploited by attackers: 1) it defines different criteria for frame validity between senders and receivers, allowing attackers to exploit this discrepancy (\IF). 2) the types of errors and their impact, namely halting transmission and raising an error frame, as well as 3) the presence of behaviorally distinct error states, are exploited in attacks that inject errors to push victims into the attacker's desired error state and in attacks that exploit the different behavior of the error-passive state (\Cc, \Ie, \Ij, \Ac, and \Ad). 4) it leaves certain cases of frame non-conformity outside its covered error conditions, allowing attackers to transmit malformed frames without provoking a uniform standard response (\Ig). Paradoxically, some of these properties are used for defense. Specifically, without {frame destruction} (\DIe) as an impact of an active error, {impersonation and replay} would be exposed as gaps (Tbl. \ref{tbl:bigtable}). Similarly, thanks to {error-states and error-types}, fast {layer-II node-suspension} (\DAd) is an effective {bus-access denial} defense, without which it would be exposed as a {by-the-rules} gap.

\shortsectionEmph{$\mathcal{R}$5: Physical Signaling (I).} Several characteristics of CAN physical signaling are exploitable by attackers ranging from {by-the-rules} attackers to {down-to-earthers}: 1) despite differential voltage and twisted wires, CAN is still vulnerable to emitting EMI (\Ce) due to lack of shielding and relatively high switching speed. 2) minor electrical differences among nodes using the bus as a shared electrical medium allow {hardware and topology profiling} (\Cd) and {physical fingerprint distortion} (\Ij). 3) using {push-pull signaling} to drive dominant bits and high-impedance for recessive bits enables attacks that rely on bit-flips and pulses (\Cc, \Ie, \IF, \Ii, \Ac, \Ad, \Ag). 4) dynamic edge synchronization allows {bit-bangers} manipulating edges to disrupt synchronization among nodes (\Af). These features are {inherent} but {not unique}. As will be shown in Sec. \ref{sec:investigat}, many other protocols' physical signaling share them.\looseness-1

\shortsectionEmph{$\mathcal{R}$6: CAN Bit Sampling (IU).} To translate voltage levels into bit values, CAN nodes sense the voltage at specific times called {sample points} whose timing is {configurable}, not the same across nodes, and happens {once per bit}. Exploiting discrepancies between node sample points allows attackers to insert signal transitions between them to send messages that have differing interpretations (\Ih). This is {inherent} and {unique} as other protocols do not combine these sample-point qualities, particularly configurability.\looseness-1

\shortsectionEmph{$\mathcal{R}$7: Vulnerable Higher-Layer Protocols and Implementation.} Higher-layer protocols and their implementations can introduce exploitable logic and software weaknesses. Vulnerable features, such as request-response exchanges, session mechanisms, and functionality-disabling commands, can be abused by {by-the-rules} attackers to launch {ECU functionality disabling} (\Aa), and {sybil exhaustion} (\Ab) attacks. At the implementation level, software flaws such as memory-corruption bugs and weaknesses in diagnostic-security handling can be exploited for privileged information exposure and internal-ECU-manipulation attacks (\Cb, \Ib, and \Ic). This is {not inherent} as the CAN standard does not specify or mandate any implementations or higher-layer protocols. It is also {not unique}, as these protocols and implementations could be used with other communication media.\looseness-1

\shortsectionEmph{$\mathcal{R}$8: Layer-II Rule Violation.} The ability of {bit-bangers} to bypass CAN's controller layer rules, placing and reading arbitrary signals on the bus, underlies all {bit-banger} attacks (\IF-\Ii, and \Af-\Ag). It also underlies the {bit-banger} variants of several attacks that could be launched using different methods for {by-the-rules} attackers (\Cc, \Ie, \Ij, and \Ac-\Ae). By definition, this is {not inherent} to CAN as it does not stem from its rules but from the violation of them, jeopardizing other parts of the standard. It is either an ECU design problem allowing for the bypass or manipulation of the CAN controller, or an access problem, allowing physical installation of non-CAN-conforming devices (Sec. \ref{subsubsec:Dattacker}).\looseness-1

\shortsectionEmph{$\mathcal{R}$9: Layer-I Rule Bypass.} {Down-to-earthers}' ability to bypass CAN’s physical-layer rules through privileged physical access and specialized equipment allows them to inject or collect physical signal measurements in ways CAN was not designed for. This, along with other root causes, underlies {down-to-earther} attacks such as {hardware and topology profiling} (\Cd) and {capturing electromagnetic emissions} (\Ce).

\subsection{Defense Gap Root Causes ($\mathcal{DR}$s)}\label{subsec:defense_rc}

\shortsectionEmph{$\mathcal{DR}$1: Cryptography Challenges (I).} Cryptography on CAN faces three challenges. First, \textit{performance}: commercial ECUs often have limited processing budgets, making cryptographic operations costly if applied to all outgoing and incoming messages. Second, \textit{message length}: CAN’s 64-bit maximum payload complicates carrying authentication data. For example, AUTOSAR suggests using 24 authentication bits. If a message carries more than 40 bits of data, little room remains for authentication. If it carries less, authentication data can significantly increase message length and busload. The same length constraint also makes conventional block-cipher encryption cumbersome, since block sizes often exceed the message size. Third, and more fundamentally, CAN faces an {inherent} but not {unique} \textit{shared-media key-management dilemma}. On a broadcast medium, group keys \cite{woo2014practical, pfeiffer2017implementing, lu2019leap} preserve efficiency and can protect against {external attackers} (i.e., an ECU without the group key). However, communication is exposed to {internal attackers}: a compromised ECU that holds the group key can decrypt all traffic and impersonate any group member. Asymmetric cryptography for general communication is often impractical on resource-constrained ECUs. Pairwise keys between each sender-receiver pair \cite{pfeiffer2017implementing, lu2019leap, hartkopp2012macan} reduce that exposure, but do not fit the broadcast nature of CAN and may impose substantial communication overhead, since the same message may need to be sent multiple times under different keys. While workarounds have been proposed to address {performance} and {message length}, little has been proposed to solve this {shared-media key-management dilemma}, the main culprit behind the {internal attacker} issue. For key establishment, some works rely on a central server \cite{pfeiffer2017implementing, lu2019leap, revo2021, expiry2019}, while others propose mechanisms such as identity-based signatures \cite{groza2019identity}, implicit certificates \cite{expiry2019}, or schemes based on transwire or controller-layer characteristics \cite{mueller2015plug, jain2016physical, pfeiffer2017implementing}. Yet many of these approaches still assume pre-shared secrets, which become particularly dangerous when combined with group-key settings, since compromising one node jeopardizes the entire group. Finally, although some key-exchange protocols could in principle support revocation by re-running the exchange, few CAN works directly address internal-attacker revocation \cite{revo2021, wu2019research}, and reported revocation times remain comparable to key-exchange or long-term-key update times.\looseness-1

Overall, fixing the {performance} and {message length} challenges mainly improves bus or ECU efficiency, whereas addressing the {shared-media key-management dilemma} fills the {communication surveillance gap} (\Ca) and provides effective defenses against {bypassing diagnostic authentication} (\Ic)~and {impersonation and replay} (\Id).

\shortsectionEmph{$\mathcal{DR}$2: Costly Attacker Signal Filtering and Disabling (I).} On a broadcast bus, availability attacks aimed at disrupting communication (\Ac-\Af) can only be stopped by filtering the attacker's signals or disabling the attacker node, so other nodes have room to communicate. Similarly, attacks that rely on injecting malicious signals into specific parts of a message (\Cc, \IF) can only be prevented this way. The lack of a cheap and effective attacker filtering and disabling approach is the fundamental reason for the {bit-banger} gaps of \Cc, \IF, \Ac-\Af, where defenses cannot simultaneously achieve prevention for all attacker models and good infrastructure compatibility. Layer-II node suspension (\DAd) is a cheap form of node disabling but works only against {by-the-rule}s attackers, leaving the bus vulnerable to other attacker models. Addressing stronger attackers by adding guardians (\DIk) or fragmentation mechanisms (\DAe) requires expensive infrastructure changes. Combined with the existing detection approaches, addressing this challenge closes these gaps and provides an effective defense against integrity and availability attacks against the transwire layer (\Ij, \Ag).\looseness-1

\shortsectionEmph{$\mathcal{DR}$3: Difficult Environment Understanding.} Determining whether a message is malicious may be infeasible without understanding the bigger system (\DIb). For example, if an attacker compromises the brake ECU and sends fake brake values without impersonation, the only way to detect this fabrication is to understand how these values relate to the bigger system (e.g., the vehicle), which may involve physical modeling. This $\mathcal{DR}$ is not inherent to CAN but to its environment (e.g., vehicle) and could be solved for open or less complex environments (e.g., an elevator). It is also {not unique}, as it will be present in other communication media operating within the same environment. Addressing this root-cause closes the {fake data fabrication} gap (\Ia) and effectively detects tampering with in-flight frames (\Ie, and \Ii), which could be turned into prevention to provide an effective defense when combined with {frame destruction} (\DIe).\looseness-1

\subsection{Observations and Takeaways}
\shortsectionEmph{On CAN 2.0 and Root Causes.} Most root causes are not unique to CAN. Specifically, only vulnerable \textit{error handling} ($\mathcal{R}$4) and \textit{bit sampling} ($\mathcal{R}$6) are unique to the CAN family. Interestingly, no defense-gap root cause is unique to CAN.
\begin{tcolorbox}[
colback=blue!5,
colframe=black!40,
fonttitle=\bfseries,
boxsep=0pt,
left=2pt,
right=2pt,
top=2pt,
bottom=2pt,
before skip=4pt,
after skip=4pt,
]
CAN fundamental security problems stem from shared single-channel media properties, vehicular constraints, and ECU implementation weaknesses that could recur in any shared-media system under similar conditions.
\end{tcolorbox}
\shortsectionEmph{On CAN 2.0 and Attacks and Gaps.} An attack or a gap usually depends on multiple root causes. If any required root cause is not inherent, the attack is not inherent. Conversely, if any of the root causes underlying an attack or a gap is unique, it is unique. Thus, although 6/9 attack root causes are inherent, only 9/22 attacks are inherent, 8/22 attacks are unique, and only 5/22 are both inherent and unique. Relative to gaps, only one is inherent, and it is inherent to shared-media communication, not CAN. The 4 unique gaps are unique only because their underlying attacks rely on CAN-unique root causes, not because the absence of effective defense is unique. Most importantly, no gap is both inherent and unique.

\section{Transferability Analysis}
\label{sec:investigat}

In this section, we assess whether CAN security problems are likely to persist in other in-vehicle buses (IVBs). Building on the root causes identified in Sec.~\ref{A:Analysis}, we investigate whether the same root causes appear in alternative IVBs and extrapolate whether the corresponding CAN attacks and gaps are likely to carry over. The premise is that if an IVB inherits the root-cause pattern of an attack or gap, then the attack or gap is likely to transfer. Our investigation is therefore a targeted standard review centered on the identified root causes, rather than a broad security review of each IVB. Where standard review is insufficient to determine the presence of a root cause, we may use formal verification, as with the optional standardized link-layer security mechanisms CANsec and MACsec.

Tbl.~\ref{tbl:root_cause} summarizes the transferability analysis. IVB rows report root-cause presence, where ($\CIRCLE$) means largely present, ($\LEFTcircle$) somewhat present, and ($\Circle$) largely absent. IVB columns report the resulting attack/gap transferability, where ($\CIRCLE$) means likely, ($\Circle$) unlikely, and ($-$) unresolved. Excluded attacks and root causes are shown in gray.

\shortsectionBf{Emerging IVB Selection.} Although point-to-point \cite{MIPI, ASA} and switched communication technologies \cite{AEth100, AEth1000} are appearing in vehicles, they require substantially more wiring. The wiring harness is already among the vehicle’s most costly components and its third heaviest after the engine and chassis \cite{Wiring2, Wiring}. We therefore expect wiring cost and weight to confine such technologies to essential use cases, with buses continuing to dominate in-vehicle communication. Among the many existing choices, we focus on three representative in-vehicle bus technologies that are strong candidates to replace classic CAN \cite{Bosch2022CANXL,Schaefer2023ArtNetworking}: CAN FD, CAN XL, and Automotive Ethernet (10BASE-T1S). We select them because manufacturers are most likely to adopt successors built on already established automotive communication ecosystems, namely CAN and Ethernet. To separate security problems rooted in bus protocols from those rooted in the surrounding system, we further consider a hypothetical ideal bus that provides secure logical channelization with confidentiality, integrity, and availability.

\shortsectionBf{Investigation Methodology.} To ensure a fair cross-IVB comparison, we focus only on weaknesses rooted in the standard. Accordingly, we adopt a by-the-rules attacker model and exclude attacker models that depend on violating, bypassing, or otherwise stepping outside the standard rules. We also assume that vehicle higher-layer protocols running on top of CAN could run on top of any of the considered IVBs. Under these comparison rules, $\mathcal{R}$1 and $\mathcal{R}$7 are satisfied for all IVBs, while $\mathcal{R}$8 and $\mathcal{R}$9, together with the attacks that rely on them, are excluded from the comparison.

\subsection{Transferability Analysis}

\mysubsub{CAN FD} Reviewing the standard \cite{bosch2012can,ISO1,ISO2}, CAN FD is a shared single-channel broadcast bus and thus shares the same {universal shared media properties} ($\mathcal{R}$2). It uses the same priority system as CAN 2.0 and its priority is settable for {by-the-rules} attackers ($\mathcal{R}$3). Despite slight modifications, its error-handling mechanism retains the same {different definitions of frame validity}, {error types and impacts}, {behaviorally distinct states}, and {corner cases} ($\mathcal{R}$4). It uses the same physical signaling with {higher switching speeds} than CAN 2.0 and without stipulating shielding as a requirement. Similar to CAN 2.0, only dominant bits are driven by push-pull. Nodes perform {dynamic edge synchronization} ($\mathcal{R}$5). Bits are sampled by reading the voltage level {once at configurable sample points} ($\mathcal{R}$6). It takes no measures to allow {selective attacker signal filtering} ($\mathcal{DR}$2). While its longer payload alleviates a great part of the {performance and length} challenges outlined in $\mathcal{DR}$1, it is still subject to the same {shared media key-management dilemma}.\looseness-1

\mysubsub{CAN XL}\label{subsub:CXL} Reviewing the standard \cite{ISO1, ISO2}, CAN XL is a shared single-channel broadcast bus and shares the same {universal shared media properties ($\mathcal{R}$2)}, enabling {communication surveillance} (\Ca) and {impersonation and replay} (\Id). It uses a similar priority system but conventional ID functionalities are split into a {Priority ID} and an {Acceptance Field}, both settable by {by-the-rules} attackers ($\mathcal{R}$3), making {bus-access denial} (\Ae) likely.\looseness-1

Error handling could be enabled or disabled by {by-the-rules} attackers. \textbf{When enabled}, it is similar to CAN 2.0, suggesting equivalents of all error-injection-based attacks (\Ac, \Ad, \Cc, \Ie, \Ij) are possible. \textbf{When disabled}, considerable changes exist, including removing {behaviorally distinct error states} ($\mathcal{R}$4). This likely avoids {error state manipulation} (\Ad). However, this also removes {node suspension} (\DAd) as a possible defense, as a {by-the-rules} attacker can disable its own error handling so it is never suspended into the bus-off state, regardless of the victim's setting. Since {node suspension} is an effective defense against many attacks, and the only effective defense against {bus-access denial} (\Ae), it is possibly exposed as a gap for both error handling settings. Both 0s and 1s are driven using push-pull. This still makes equivalents of {error injection} likely. However, the outcome of a 0-1 collision will be less predictable than CAN 2.0, where it would always be 0. Therefore, the presence of {frame hijacking} (\Ie) and {network reconnaissance} (\Cc) equivalents cannot be determined and needs further research. 

It supports {higher switching speeds} and, similar to CAN 2.0,  does not mandate shielding ($\mathcal{R}$5), uses {dynamic edge synchronization}, with bits read {once at configurable sample points} ($\mathcal{R}$6). It is compatible with CANsec \cite{CANsec} and its length is significantly longer, alleviating {performance and length} challenges. To assess if it addresses the {internal attacker} issue ($\mathcal{DR}$1), we formally verified CANsec (App. \ref{inv:CANsec}). Our verification shows that as the current in-development version of CANsec stands, it provides authenticity, freshness, and confidentiality only against external attackers but not {internal attackers} ($\mathcal{DR}1$) and hence does not close any open defense gaps.

\mysubsub{10BASE-T1S} Reviewing the standard \cite{AEth10B}, 10BASE-T1S is a shared half-duplex broadcast bus, with the same {universal shared media properties} ($\mathcal{R}$2), making equivalents of {communication surveillance} (\Ca), and {impersonation and replay} (\Id) possible. Priority is managed by Physical Layer Collision Avoidance (PLCA). Each node's priority is determined by its local ID. Similar to CAN, it is settable for {by-the-rules} attackers ($\mathcal{R}$3),  making {bus-access denial} (\Ae) possible. 

It has different error handling rules and no error states, likely avoiding {error state manipulation} (\Ad). However, it lacks features that could be used as effective defenses, such as frame destruction (\DIe) or node suspension (\DAd), which in turn is likely to expose {impersonation and replay} (\Id) and {bus access denial} (\Ae) as gaps. {Network reconnaissance} (\Cc) is {likely}, not using errors, but using inherent properties (e.g., node ID). Research on whether it harbors properties allowing equivalents of {error injection} (\Ac), {frame hijacking} (\Ie), and {fingerprint distortion} (\Ij) is needed.

Similar to CAN, it uses differential signaling with {dynamic edge synchronization}, does not stipulate shielding, and supports {higher switching speeds} than CAN 2.0. It uses differential Manchester encoding, where both 0s and 1s are driven with push-pull. This makes outcomes of 0/1 collisions unpredictable, unlike CAN, where it is always 0 ($\mathcal{R}$5). Unlike CAN, receivers interpret bits by reading voltage transitions at fixed start and middle of bits, without configurable sample points ($\mathcal{R}$6).\looseness-1

Like CAN, it takes no measures to facilitate {selective attacker signal filtering} ($\mathcal{DR}$2). It is natively compatible with MACsec, and its maximum length is significantly longer, alleviating the {performance and length} challenges. To assess if it addresses the {internal attacker} issue ($\mathcal{DR}$1), we formally verified MACsec (App. \ref{inv:CANsec}). Our verification shows that as the 10BASE-T1S version of MACsec stands, it provides authenticity, freshness, and confidentiality only against external attackers but not {internal attackers} ($\mathcal{DR}1$) and hence does not close any open defense gaps.

\mysubsub{\MR{6}Hypothetical Ideal Bus} To distinguish between security problems that can, in principle, be eliminated by bus protocol design and those that can only be addressed by securing the system surrounding the bus, we assume the presence of an ideal bus. Note these are not meant to be practical recommendations on how a more secure bus should be designed, but establish an upper bound on the security guarantees achievable by bus protocol design alone. The bus offers secure logical channelization. Each channel consists of one authorized transmitter, one or more authorized receivers, and a defined set of messages. Transmission from different channels may occur simultaneously on the same physical medium (using Code Division Multiple Access CDMA or other methods). Nodes outside a channel cannot observe messages inside a channel or learn their contents (for example, using VLAN-like technologies). This eliminates $\mathcal{R}$2. A message can only be transmitted by the authorized transmitter and its priority cannot be tampered with (eliminating $\mathcal{R}$3). Nodes' transmission opportunities are guaranteed and no single node can deny all other nodes from bus access (using scheduling, or other methods). All these rules are enforced by protocol controllers attached to each node. The controller cannot be bypassed or tampered with by a by-the-rules attacker, and bus access can be revoked from compromised nodes (addressing $\mathcal{DR}$2). Nodes in a channel share a channel-specific key. Messages are protected using fast authenticated encryption. If a compromised channel member is detected, a fast revocation mechanism removes it from the channel and securely distributes new keys to the remaining members (eliminating $\mathcal{DR}$1). CAN-specific error handling ($\mathcal{R}$4) and bit sampling ($\mathcal{R}$6) properties are eliminated (by adopting mechanisms different from CAN). $\mathcal{R}$5 is partially addressed by physical layer design. The bus may avoid push-pull signaling, mandate shielding to reduce EMI, and use a shared clock to avoid dynamic edge synchronization. However, transmitters will still exhibit distinguishable physical signal characteristics due to hardware differences.

\begin{figure}[t]
  \centering
  \includegraphics{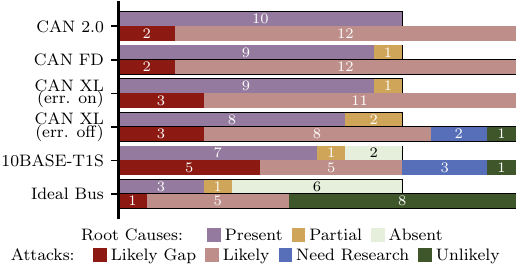}
  \caption{Transferability of CAN 2.0 security problems to emerging IVBs.}
  \label{fig:ivb}
\end{figure}

Even with such an ideal hypothetical bus, $\mathcal{R}$1, $\mathcal{R}$7, and $\mathcal{DR}$3 will remain, reproducing attacks \Cb, \Ia-\Ic, \Aa-\Ab~and the fake data fabrication gap. These causes are rooted in the system surrounding the bus, not the bus itself. They will reproduce no matter how perfect the bus is. Specifically, appropriate control is needed to restrict physical access or remote exploitation of vulnerabilities in bus-connected devices. Otherwise, an attacker can still access an ideal secure bus ($\mathcal{R}$1). Insecure features and implementations of higher layer protocols ($\mathcal{R}$7), and more accurately modeling the cyber-physical environment ($\mathcal{DR}$3), are outside the scope of what the physical and data link layer specifications of a bus can secure.

\subsection{Observations and Takeaways}

\shortsectionEmph{On IVBs and Root Causes.} Fig. \ref{fig:ivb} summarizes the results of the transferability analysis. As shown, \textbf{CAN FD} shares all compared CAN root causes, making all considered attacks and gaps likely transferable. \textbf{CAN XL} has two error-handling settings. When enabled, it fully shares 9 root causes, and partially shares 1. As such, 14 attacks and 2 gaps under comparison are likely transferable, and 1 additional attack is likely exposed as a gap. When disabled, it fully shares 8 root causes, and partially shares 2. Consequently, 11 attacks and both gaps are likely transferable. 1 attack is unlikely to transfer, and 2 require further research to determine. 1 additional attack is likely exposed as a gap. \textbf{10BASE-T1S} fully shares 7 root causes, partially shares 1, and avoids 2. As such, 10 attacks are likely transferable, 3 need further research, and 1 is unlikely to transfer. Both gaps are likely transferable with 3 additional attacks likely exposed as gaps. \MR{6}Finally, an \textbf{ideal secure bus} fully shares 3 root causes, partially shares 1, and avoids 6. 8 attacks may be avoided and 6 remain, where 1 remains as a gap. This shows that certain problems are rooted in the system surrounding the bus and will reproduce no matter how perfect the bus is unless a new approach to vehicle security is taken, where the entire system is considered as one whole. 

 \begin{tcolorbox}[
 colback=blue!5,
  colframe=black!40,
  fonttitle=\bfseries,
  boxsep=0pt,
  left=2pt,
  right=2pt,
  top=2pt,
  bottom=2pt,
  before skip=4pt,
  after skip=4pt,
]
\begin{itemize}[leftmargin=*]

\item Emerging IVBs do not solve CAN's fundamental security problems. Most CAN security problems likely transfer.
\item CAN is more securable than perceived. Each examined IVB inherits at least as many transferable attacks with defense gaps as CAN, before accounting for IVB-specific risks.\looseness-1
\item Certain CAN security problems can only be solved by securing the system surrounding the bus.
\end{itemize}
\end{tcolorbox}

\shortsectionEmph{\MR{7}Limitation.} The findings of the transferability analysis were only indicative. The attacks identified as likely transferable were not validated on the respective buses. Such validation could be performed by formally analyzing parts of each bus protocol to determine attack viability, or empirically by setting up a testbed. Of the 14 compared attacks, 5 relied on higher-layer protocols. These attacks would need to be evaluated not only on the respective buses, but with the higher-layer protocols also set up on the ECUs. For CAN variations (CAN FD and CAN XL), most techniques used in CAN 2.0 could be directly utilized. However, for 10BASE-T1S, some techniques would need to be adapted to the specifics of the protocol. For example, instead of achieving bus access denial by faking a high-priority CAN ID, an attacker may need to fake the node ID. Across all buses, attack success could then be measured individually for each attack using the same or adapted metrics the attack authors used (e.g., swiftness and suppression rates for error-state manipulation/bus-off attacks, message loss and delay for bus-access denial attacks, etc.).



\section{Future Directions}

\label{sec:directions}



\begin{figure}[t]
  \centering
  \includegraphics[width=\columnwidth]{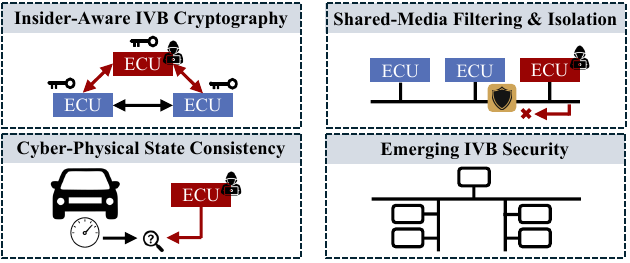}
  \caption{Research directions for future in-vehicle communication security.}
  \label{fig:direction}
\end{figure}

A major finding of our study is that many CAN security problems are rooted in causes common to single-channel shared communication media and vehicular environments, rather than in CAN-specific design choices. Accordingly, we believe the following research directions (Fig. \ref{fig:direction}) are beneficial not only for CAN 2.0, but also for any standard that may replace it.

\shortsectionBf{Insider-Aware IVB Cryptography.} Future IVB cryptography must treat compromised ECUs as first-class adversaries, not just external attackers. IVB cryptography has focused on fitting authentication and confidentiality into constrained buses: limited payloads, strict timing, low ECU compute budgets, and bus-load sensitivity. Newer IVBs reduce some barriers through longer frames, higher bandwidth, and hardware-assisted line-rate protection. However, our analysis shows that a deeper problem behind several defense gaps is shared-media key management: a compromised ECU with valid group keys can still decrypt traffic or impersonate messages. Future work should therefore focus on insider-aware IVB cryptography, fast revocation, secure rekeying, continuous attestation, and zero trust architecture.\looseness-1

\shortsectionBf{Shared-Media Filtering and Isolation.} One finding of this study is that, whatever replaces CAN 2.0, future shared-media IVBs need a practical way to contain a compromised node without taking down the whole channel ($\mathcal{DR}2$). IDSs can detect problems, and encryption may preserve confidentiality, but neither helps if the attacker can still occupy the medium, inject errors, disturb timing, force retransmissions, or manipulate physical/data-link-layer behavior. In a single-channel bus, one malicious transmitter can affect every receiver, while the network often lacks a cheap way to suppress only the malicious frame, dominant bit, error, timing disturbance, or node. Without such containment, many CAN gaps will transfer to successor buses, as shown in Sec. \ref{sec:investigat}. Existing CAN approaches, such as {ECU guardians} (\DIk) and {dynamic fragmentation} (\DAe), attempt this isolation, but their infrastructure cost limits adoption. Future work should therefore develop low-cost shared-media filtering and isolation mechanisms that selectively block messages or signals, disconnect malicious nodes, or isolate bus segments while preserving normal operation.


\shortsectionBf{Cyber-Physical State Consistency.} Many IVB attacks are not mere identity-forgery attacks, but semantic lies about the physical state. This is why fake data fabrication remains a difficult gap to close in CAN and the other IVBs we investigated: a forged speed or brake value may be well-formatted, correctly timed, and even authenticated if sent by a compromised brake ECU. Detecting such attacks requires determining whether a message is physically and operationally plausible in the broader cyber-physical context. Future research should therefore move toward cyber-physical state consistency checking: determining whether a bus message is plausible given driver input, sensor readings, actuator behavior, ECU state, timing, control logic, and vehicle dynamics. Research should focus on capturing both expected message-level behavior and the physical consequences those messages should produce across ECUs and vehicle subsystems. Digital twins could provide the foundation for this direction. However, challenges regarding proprietary designs, incomplete knowledge, runtime uncertainty, and vehicle-specific behavior are expected to arise.

\shortsectionBf{Emerging IVB Security.} One finding of this study is that newer IVBs likely inherit many of CAN's security problems. Higher bandwidth, longer payloads, revised frame formats, and optional security mechanisms can improve important constraints, but they do not remove issues arising from shared access, broadcast communication, priority mechanisms, physical signaling, or compromised insiders. However, our investigation only establishes root-cause presence and uses it to infer the likelihood of transferability. Future research should therefore validate these indications by determining whether the corresponding attacks and gaps are practically exploitable, how they unfold in each IVB, and what countermeasures can stop them. More importantly, research should not be limited to CAN-transferable problems: each emerging IVB may introduce protocol-specific vulnerabilities through its own frame format, media-access rules, encoding, error behavior, and security extensions. A recent formal analysis of CAN XL validates several of our transferability predictions, showing that CAN XL retains known CAN 2.0 MAC-layer vulnerabilities and their corresponding attacks, while also uncovering new CAN XL-specific vulnerabilities beyond our comparison \cite{tang2026formal}. This work is time-sensitive: once an IVB becomes widely deployed, structural weaknesses become difficult and expensive to correct. \looseness-1

\section{Conclusions}
In this paper, we addressed a critical gap in CAN security literature: the lack of a coherent systematization framework. Through an extensive literature survey, we introduced a comprehensive taxonomy and assessment models to articulate attacker privileges, the attacks they enable, their targets, replicability, defense efficacy, and impact. Leveraging this framework, we highlighted where the literature is concentrated or lacking and identified CAN’s defense gaps. We then analyzed the root causes of CAN insecurity to determine whether they are inherent or unique to CAN, and conducted a follow-up investigation to assess whether future IVB technologies decisively resolve conventional CAN security challenges. Our findings challenge common perceptions: CAN is more securable than often assumed, most of its fundamental security problems stem from shared single-channel communication properties or vehicular constraints, and most of these problems are transferable to other IVBs. Finally, we concluded that securing CAN and future IVBs requires addressing shared root causes more than CAN-specific causes, and outlined four key research directions to secure IVB systems down the road.

\section{Acknowledgments}
We thank the anonymous reviewers for their comments and suggestions. This work was supported in part by the Office of Naval Research (ONR) under Grant N00014-18-1-2674, the National Science Foundation (NSF) under Grants CNS-2145744 and CNS-214464, and the National Center for Transportation Cybersecurity and Resiliency (TraCR) under Award Number 2589-211-2026310. Any opinions, findings, conclusions, and recommendations expressed in this material are those of the authors and do not necessarily reflect the views of the ONR, NSF, or TraCR, and the U.S. Government assumes no liability for the contents or use thereof.
\section{\MR{4}Ethics Considerations}
\label{section: ethical}

 As a systematization of knowledge paper, this study did not involve manipulating real-world systems. Our formal verification explored the protection boundary of CANsec/MACsec and whether it extends beyond current cryptographic CAN solutions. It did not identify previously unknown protocol vulnerabilities. Nevertheless, we have been in contact with CAN in Automation (CiA). They approved our request to access the in-development CANsec draft standard and conduct this research, and have been informed of our results. They asked to emphasize that CANsec and CKA are under development, which we have done. All works referenced in this paper are publicly accessible. No private or confidential information was used.

{
\bibliographystyle{IEEEtran}
\bibliography{reference}
}

\appendices

\section{\MR{4}CANsec / MACsec Formal Verification}\label{inv:CANsec}
As a supplementary check within the IVB root-cause analysis, we examine whether CANsec and automotive MACsec close the internal-attacker gap ($\mathcal{DR}$1). In their automotive configurations, both organize nodes into Connectivity Associations (CAs) whose members share a long-term symmetric Connectivity Association Key (CAK) to authenticate then establish short-term  Secure Association Keys (SAK) to protect communication. However, a shared CAK alone does not determine insider protection because the protocols and their key-agreement mechanisms could still introduce per-sender or pairwise keys, node-specific derivation, or compromised-member exclusion and rekeying. We therefore model the protocols and the key-management configurations specified or currently proposed for automotive use and verify their security properties against both external and internal attackers. We find that none of the modeled configurations introduces key or privilege separation among members of the same CA, or other mechanisms that could close $\mathcal{DR}$1. Since MACsec supports other key-management configurations and CANsec remains under development, we additionally test whether idealized revocation used in conjunction with these protocols could close the gap. The verification code and additional details are available at~\cite{code}.

\begin{figure}[t]
\centering
\includegraphics{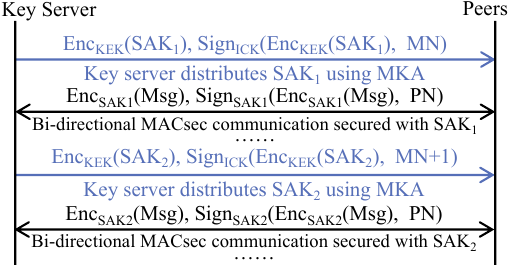}
\caption{MACsec and MKA.}
\label{fig:macsec}
\end{figure}

\shortsectionBf{MACsec.} MACsec is a link-layer security add-on compatible with all Ethernet variants that provides message authenticity, freshness, and confidentiality~\cite{MACsec}. Nodes form a CA. MKA~\cite{mka} authenticates CA members that hold a CAK and includes a key-server role. As shown in Fig.~\ref{fig:macsec}, the shared CAK is used to derive a Key Encrypting Key (KEK) and Integrity Check Key (ICK). The key server generates SAKs, encrypts them using the KEK, and authenticates their distribution using the ICK. CA members derive the same KEK and ICK from the shared CAK, obtain the distributed SAKs, and use them to protect regular MACsec communication.

\begin{figure}[t]
\centering
\includegraphics{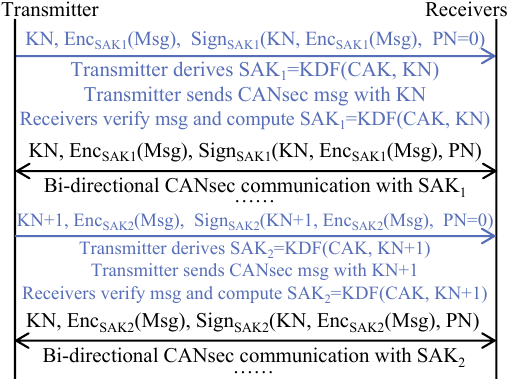}
\caption{CANsec and IKA.}
\label{fig:cansec}
\end{figure}

\shortsectionBf{CANsec.} CANsec is an \textit{in-development} link-layer security add-on for CAN XL modeled after MACsec~\cite{CANsec}. Its current proposal similarly uses CAs, CAKs, and SAKs, with CANsec Key Agreement (CKA) establishing SAKs. CKA is \textit{still in early development}, but two approaches are currently being considered: 1) an approach modeled after MKA and 2) In-line Key Agreement (IKA)~\cite{zeh2025secure}. In IKA, SAKs are derived using the CAK, an incrementing Key Number (KN), and a Key Derivation Function (KDF), as shown in Fig.~\ref{fig:cansec}.

\shortsectionBf{Assumptions and Adversary Model.} We use ProVerif~\cite{blanchet2016modeling} to model the protocols and attackers. While MKA supports multiple configurations and optional features, including flexibility in how the CAK itself is established and how members are authenticated, relevant automotive MKA specifications and AUTOSAR~\cite{automotive_mka,autosarmka} assume a pre-shared CAK with a fixed key server. We make the same assumption. For CANsec, both the currently proposed MKA-based CKA and IKA assume a pre-shared CAK. We therefore model both with a pre-shared CAK. We assume a by-the-rules attacker following the Dolev--Yao model with two variations: an external attacker who is not a member of the CA, and an internal attacker who controls a compromised member ECU and has access to its CAK.

\shortsectionBf{Verified Properties.} We verify the following properties against both attackers:
\begin{itemize}[itemsep=3pt]
\item[$\mathcal{P}_{a}$:] An attacker cannot cause an honest CA member to accept spoofed CA messages or SAK distribution as originating from another honest member.
\item[$\mathcal{P}_{f}$:] An attacker cannot cause an honest CA member to accept replayed CA messages or SAK distribution as fresh.
\item[$\mathcal{P}_{c}$:] An attacker cannot learn protected CA communication exchanged between other members.
\end{itemize}

\shortsectionBf{Results.} In MKA, CA members derive the same KEK and ICK from the shared CAK and can therefore obtain the SAKs distributed by the key server. The MKA-based CKA approach follows the same general structure. IKA establishes SAKs differently, but every CA member possessing the shared CAK can derive them using the CAK and KN. Thus, despite their key-management and key-establishment machinery, neither protocol introduces key or privilege separation among members of the same CA, or any other mechanisms that protect against internal attackers. Accordingly, $\mathcal{P}_{a}$, $\mathcal{P}_{f}$, and $\mathcal{P}_{c}$ hold against the external attacker but not the internal attacker. Neither automotive MACsec nor the current CANsec configurations therefore close $\mathcal{DR}$1.

\shortsectionBf{Hypothetical Revocation Experiment.} The results above identify the inability to exclude a compromised member and rekey the remaining group as a key limitation of the considered configurations. We therefore test whether adding this capability closes the internal-attacker gap. We model an idealized revocation mechanism that immediately removes the compromised member from the CA and securely distributes a new CAK and SAKs to the remaining members. After revocation, $\mathcal{P}_{a}$, $\mathcal{P}_{f}$, and $\mathcal{P}_{c}$ hold against both external and internal attackers, closing $\mathcal{DR}$1. Our goal is not to propose a revocation protocol, but to test whether this capability is sufficient to address the identified limitation and motivate future research into practical mechanisms that provide it.

\shortsectionBf{Limitations.} Our conclusions apply only to the configurations modeled above. MACsec supports additional authentication and key-management mechanisms that are not incorporated into the relevant automotive profiles and are outside our scope. Our results therefore should not be generalized to MACsec configurations using such mechanisms. CANsec and CKA remain under development, and our conclusions reflect the current proposal. They may differ for the eventual standardized design. The revocation experiment tests only whether ideal, immediate revocation is sufficient to close $\mathcal{DR}$1 and does not propose or evaluate a practical revocation protocol.

\end{document}